# A Local-Realistic Model of Quantum Mechanics Based on a Discrete Spacetime (Extended version)

Antonio Sciarretta

34 rue du Château, Rueil Malmaison, France

Tel.: +39.147525734, E-mail: asciatopo@gmail.com

**Abstract**

*This paper presents a realistic, stochastic, and local model that reproduces nonrelativistic quantum mechanics (QM) results without using its mathematical formulation. The proposed model only uses integer-valued quantities and operations on probabilities, in particular assuming a discrete spacetime under the form of a Euclidean lattice. Individual (spinless) particle trajectories are described as random walks. Transition probabilities are simple functions of a few quantities that are either randomly associated to the particles during their preparation, or stored in the lattice nodes they visit during the walk. QM predictions are retrieved as probability distributions of similarly-prepared ensembles of particles. The scenarios considered to assess the model comprise of free particle, constant external force, harmonic oscillator, particle in a box, the Delta potential, particle on a ring, particle on a sphere and include quantization of energy levels and angular momentum.*

## 1 Introduction

### 1.1 Motivation

One of the most intriguing aspects of quantum mechanics (QM) is that individual particles that do not interact with each other can exhibit interference patterns when statistical distributions of measured events are considered. Feynman stated that the phenomenon of electron diffraction by a double-slit or double-source preparation is "impossible, absolutely impossible to explain in any classical way, and has in it the heart of quantum mechanics. In reality it contains the only mystery" [1]. Many of the typical quantum phenomena involving nonrelativistic spinless particles (among which discrete levels of energy and angular momentum, the uncertainty principle, particle-wave duality, tunnelling) are indeed related with wave-like behaviour and wave superposition in particular.

Quantum theory postulates that it is fundamentally impossible to go beyond the description of interference patterns in terms of probability distributions and does not provide any mechanism to describe the individual events that perhaps contribute to the observed statistical averages. As such, the interpretation of QM formalism is still the subject of debate. Most physicists embrace the orthodox view that the description offered by quantum mechanics is complete and that only probability distributions can be described. In contrast with this attitude, "realistic" interpretations have been sought since the early quantum history, i.e., ontological models based on real physical states not necessarily completely described by QM, but objective and independent of the observer [2, 3, 4, 5]. In realistic theories, at a given time, particles have definite values for any possible observable (before that a measurement is made). In particular, particle trajectories are assumed to exist.

Perhaps the best-known realistic theory, the De Broglie–Bohm mechanics [6, 7] describes individual particles as following well-defined trajectories. For a given initial position, each trajectory is deterministic and regulated by a mechanism involving a guiding wave (or a quantum potential). The emergence of the stochastic behaviour leading to the collective patterns observed is still a matter of debate, however, the general consensus is that probability represents the uncertainty of the initial state of the particle rather than a possible randomness in the trajectories.

The Bohmian mechanics explicitly appeals to a nonlocal mechanism, with the guiding wave or the quantum potential instantaneously influencing the particle trajectory far away. The resulting "nonlocal realism" is generally accepted as it violates Bell's inequality. However, local mechanisms are not universally dismissed, as the accuracy of Bell test experiments and the conclusions drawn from them are still a subject of profound debate [8-11].

We aim at exploring the possibility that quantum-like behaviour can be equally explained by assuming a local-realistic event-based behaviour, that is, at the same time, complying with the principle of realism and that of locality. Additionally, we interpret the emergence of QM probability distributions as a result of a fundamental randomness. In summary, we assume that each individual particle has a definite trajectory (R) that is stochastic (S), and subject to purely local interactions (L). Of course, we search for an algorithm that produces individual events, with frequencies agreeing with QM probability distribution, without referring the algorithm to the distribution itself.

To our best knowledge, this program has not been fully exploited as yet.

Features (R) and (S) are satisfied by random walks. The idea to simulate QM with random walks has its origin from the path integral formalism and the chessboard model introduced by R. Feynman [12]. Since then, many papers have appeared in the literature to retrieve the Schrödinger equation from a random walk process, including work of G. Ord [13-15] and subsequent refinements [16], as well as different yet related approaches [17]. While these approaches are able to reproduce the emergence of a Schrödinger-type equation, the Born probability rule and interference are not explained in such models.

Following E. Nelson's seminal work [18], another approach has consisted of showing the emergence of the Schrödinger equation from a stochastic equation of motion [19-24]. Such derivation, however, is founded on the assumption of reversible diffusion or competing diffusion-antidiffusion processes, leading to a key osmotic velocity that depends on the probability field it concurs in building. As such, these theories are not coping with the feature (L) above.

An event-based class of models have been recently proposed [25, 26] that embody quantum behaviour into the constitutive models of the detectors and other devices, mostly using deterministic learning machines that generate events according to QM. This approach captures all of the features (R), (S), and (L). Interference, among other quantum phenomena, is not obtained as an intrinsic characteristic of the particles' trajectories but as a result of their interaction with the experimental apparatus.

Other suggestions to make QM behaviour emerge from discrete time evolution [27] or vacuum fluctuations [2] were not pursued to the end, to our best knowledge.

In the proposed model, we interpret the results of nonrelativistic spinless quantum mechanical systems as probability distributions of similarly-prepared ensembles of particles that are emitted by one or multiple sources. At a given time, individual particles have definite values for position and momentum, among other observables, thus fulfilling the feature (R). It is further noteworthy to mention that detectors and other devices are here assumed to be simple counters and do not participate in building the interference patterns.

The stochastic behaviour (S) that is manifested by the empirical evidence of QM is explained by assuming a fundamental randomness in particles' trajectories. The emergence of QM behaviour is a consequence of the particular rules of motion chosen. The motion of individual particles and their interaction with external forces take place on a discrete space-time under the form of a lattice. Particle trajectories are asymmetric random walks, with transition probabilities being simple functions of a few quantities that are either randomly attributed to the particles during their preparation at the sources, or stored in the lattice nodes that the particle visits during the walk. The lattice-stored information is progressively built as the nodes are visited by successive emissions. This process, where particles leave a "footprint" in the lattice that is used by subsequent particles, is ultimately responsible for the QM behaviour.

Therefore the interactions between subsequent emissions fulfil the feature (L), albeit through the mediation of the lattice. It is further noteworthy to mention that in the proposed model the lattice is only the support for particle motion, not for wavefunctions or other mathematical operators [28, 29]. In a certain sense, the lattice plays the role of the "guiding wave" assumed by de Broglie-Bohm theory and evidenced by some macroscopic experiments [30]. However, the QM behaviour is not produced by a single particle interacting with the wave interface ("walker" in [30]), but it is generated by successive emissions of the same ensemble.

Being based on discrete motion, the proposed model only involves integer values for time, space, velocity, energy, etc. of a given particle. In addition, some rational-valued quantities are derived from these primary quantities. Mathematical operations on physical quantities are reduced to arithmetic operations that, in fact, ultimately reflect the fundamental probability rules. It is further noteworthy to mention that QM behaviour is not reproduced by appealing to probability cancellation, nor to other definitions of negative probabilities.

## 1.2 Introductory example: the double slit scenario

Consider a simplified double-slit apparatus, where similar particles are emitted at a certain rate from either of two sources, which are placed along, say, the $x$ direction at positions $x_0 = \pm \delta$. Emitted particles have a constant speed along the perpendicular $y$ direction, but a random initial speed along the $x$ direction. At a certain distance from the sources along the $y$ axis, a screen is placed and particle arrivals are detected. Given the constancy of the $y$-axis velocity, will assume that all particle reach the screen after the same time $T$. We shall thus consider only motion along the $x$ direction and we are interested in evaluating the frequency of arrivals $P(x)$ at the screen. The QM result is known to be $P(x) \propto 1 + 2\sqrt{P_1 P_2} \cos\left(\frac{2\pi \delta x}{T}\right)$, if $P_{\{1,2\}}$ is the probability of emission from sources 1 and 2, respectively, with $P_1 + P_2 = 1$. The proposed mechanism is capable of retrieving this result.

The first key idea is that the initial velocity component $v_{0x}$ is randomly attributed to the emitted particles within the luminal limits. The pdf of this stochastic variable is $\rho(v_{0x}) = 1/(2c)$. Secondly, it is admitted that, even in the absence of external forces, the particle's velocity varies during the motion, due to "quantum" forces. In fact, it turns out that in this double-slit situation the velocity of each particle tends to a value $v_x$ such that $v_{0x} = v_x + \sqrt{P_1 P_2}/\pi \cdot \sin(2\pi\delta v_x)$. Since $v_{0x}$ is randomly attributed, such a steady-state velocity is a stochastic variable, too, and its pdf is obtained as $\rho(v_x) = \rho(v_{0x}) dv_{0x}/dv_x = (1 + \sqrt{P_1 P_2}\cos(2\pi\delta v_x))/(2c)$, given the monotonic relationship between the two quantities. Now, the position at time $T$ is roughly $x \sim x_0 + v_x T$ for times large enough so that the velocity has stabilized. Therefore, the position pdf is found as $\rho(x) = \rho(v_x)dv_x/dx = 1 + \sqrt{P_1 P_2}\cos(2\pi\delta x/T)/(2cT)$. With that, the QM probability is retrieved. The scaling factor $1/(2cT)$ ensures that the integral of $\rho(x)$ between the maximum and minimum positions that can be reached at the screen is 1.

In case of a single source, there are no quantum forces: $v_x$ would have been equal to $v_{0x}$ and $\rho(x) = 1/(2cT)$. Clearly, the key mechanism to retrieve the QM predictions is the fact that the particle's velocity converges to a steady-state value that depends on its initial velocity and on the probability distribution of the sources. We shall illustrate in the next sections the proposed mechanism leading to that.

Continuously during their flight, the various particles leave a footprint of their passage, consisting of the distance $\ell$ they have covered so far. Particles emitted at source 1 leave a footprint $\delta + v_x t$, while for particles emitted at source 2, the footprint is $-\delta + v_x t$. Moreover, each particle is capable to sense the footprint $\lambda$ left by the particle that has previously crossed its current position at the same flight time.

There are now two possibilities. The first case is when $\lambda$ is equal to $\ell$. This happens when the flying particle leaves from source 1 and finds the footprint of a previous particle equally emitted from 1, or when the flying particle is from source 2 and finds the footprint of a previous particle equally emitted from 2. This circumstance, which we may label "a", has thus a probability $P_a = P_1^2 + P_2^2$ to happen. The second case is when $\lambda$ is different from $\ell$. This happens when a flying particle from 1 finds the footprint of a particle emitted from 2, or vice versa. Thus, in this circumstance, $|\lambda - \ell| = 2\delta$. This event "b" has probability $P_b = 2P_1 P_2$. Note that $P_a + P_b = 1$.

When event "b" occurs, a pair of entities (called "bosons" in the following) is created, one of which is attached to the flying particle and the other to the footprint. The newly created bosons replace possibly existing bosons that were already attached to the particle or to the footprint, respectively (resulting from a previous event "b"). Both bosons carry a momentum contribution. The momentum of the new particle boson $v_b$ equals the previous momentum of the footprint, $\omega$, divided by the difference $|\lambda - \ell|$, that is, $2\delta$. On the other hand, the new footprint momentum equals the particle velocity multiplied by $2\delta$. In other terms, there is an exchange of momentum information between the particle and the footprint.

Once created, the footprint momentum decays with a particular law that is approximately $\sim \exp(1/\text{time})$, unless the footprint is replaced by a new one left by another particle (another event "b"). However, the expected lifetime of a footprint is very long, since only one footprint is visited at each time by the currently flying particle, and particles are emitted at a finite rate. The footprint momentum has thus enough time to converge to a steady-state value that, as it will be shown in the appropriate section, is $\omega = \sin(\pi\overline{\omega})/\pi$, where $\overline{\omega} = 2\delta v_x$ is the momentum at the footprint creation, depending on the momentum of the particle that left the footprint.

As for the momentum of the flying particle's boson $v_b$, it is reset to the value $\sin(2\pi\delta v_x)/\pi\delta$ (see above) at each event "b". Otherwise, as long as an event "a" occurs, $v_b$ decays with a particular law that is approximately $\sim 1/\sqrt{\text{time}}$. It will be shown in the appropriate section that the average value taken by $v_b$ as a consequence of this rise-and-decay behaviour is $\sqrt{P_1 P_2} \sin(2\pi\delta v_x)/\pi\delta$, as it clearly depends on the relative probability of the events "a" and "b".

Now, the velocity of the particle is given by the difference between its initial velocity and the momentum contributed by the boson, $v_x = v_{0x} - v_b$, which yields the steady-state relationship $v_{0x} = v_x + \frac{\sqrt{P_1 P_2}}{\pi}\sin(2\pi\delta v_x)$ anticipated above.

The emergence of the square root in the formula for the average boson momentum, as well as that of the sine function in the formula for the steady-state footprint momentum can only derive from a discrete formulation of the respective decay laws. Additionally, the encounter between a flying particle and a footprint can only occur precisely if discrete space and time are assumed. These two observations motivate the existence of a discrete spacetime lattice in the model detailed in the following sections.

The paper is organized as follows. First, the one-dimensional lattice is presented (Sect. 2) alongside with the fundamental spacetime quantization. Emission of particle at sources (Sect. 3) and particle motion (Sect. 4) are subsequently described. Then Schrödinger equation is retrieved by analysing the probability density functions of ensembles of particle emissions (Sect. 5). The extension to the 2-d and 3-d cases is further presented (Sect. 6). Finally, numerical simulations (Sect. 7-10) allow a comparison between the proposed model and quantum mechanical results for several scenarios.

## 2 Lattice

The proposed model assumes a discrete spacetime. For simplicity, the description below will be limited to one dimension. The spatial values are thus restricted to integer multiples of a fundamental quantity $X$ and the temporal values are restricted to integer multiples of a quantity $T$. In the rest of the paper, except when explicitly stated, these integer values will be denoted with small Latin letters, while the corresponding physically-valued quantities will be generally denoted with a tilde.

In this model, a particle's evolution consists of the succession of discrete values $x[n]$, $t[n]$, where $n \in \mathbb{N}$ is the index that describes advance in history, here denoted as "iteration". By taking an arbitrary $x = 0$ reference, the spacetime may be thought as if it is constituted by a grid $x \in \mathbb{Q}, t \in \mathbb{N}$, or "lattice", whose nodes can be visited by the particle during its evolution.

Advance in time ("lifetime") is unidirectional and unitary, that is,

$$t[n] = t[n-1] + 1, \qquad t[n_0] = 0, \tag{1}$$

where $n_0$ is the iteration when the particle is created. Advance in space ("position") is still unitary but the particle can either advance in one of the two directions or stay at rest, according to the rule

$$x[n+1] = x[n] + v[n], \qquad x[n_0] = x_0, \tag{2}$$

where the motion is regulated by a random variable $v$ ("momentum") that can take only three values, namely, $v \in \{-1,0,1\}$.

The fundamental quantities $X$ and $T$ are related to the Compton length and time,

$$X = \frac{h}{2mc}, \qquad T = \frac{h}{2mc^2}, \tag{3}$$

where $m$ is the particle's mass, $c$ is the speed of light, and $h$ is Planck constant. Relations (3) are the same as those introduced by G. Ord and the authors of [13-17]. Here, however, a slightly different justification is proposed based on the uncertainty principle.

From the rule (2), it is clear that the particle reaches its maximum speed when $v[n] \equiv 1$, which provides, in physical units,

$$\frac{X}{T} = c \, . \tag{4}$$

On the other hand, consider the problem of observing the particle's momentum. The momentum $v$ being a random variable, its average value (later defined as propensity) cannot be directly observed but only be estimated from sample momentum. The latter is also a random variable, defined as a function of the number $N$ of observations, $\tilde{v}^{(N)}$. For $N = 1$, the possible outcomes of $\tilde{v}^{(1)}$ are (in physical units) $-c, 0, +c$. Thus the uncertainty $\Delta \tilde{v}^{(1)}$ is equal to $c$. For $N = 2$, the possible outcomes are $-c, -c/2, 0, c/2, c$, with $\Delta \tilde{v}^{(2)} = c/2$. After $N$ observations, $\Delta \tilde{v}^{(N)} = c/N$. However, observing the particle for $N$ iterations implies an uncertainty in the determination of its position as well. Since the position might change from $-NX$ to $NX$, $\Delta \tilde{x}^{(N)}$ is equal to $2NX$ (in physical units). Multiplying the uncertainty in the particle's momentum ($mv$) and that in the particle's position, one obtains

$$m \Delta \tilde{v}^{(N)} \Delta \tilde{x}^{(N)} = 2mcX = h \, , \tag{5}$$

which is compliant with Heisenberg uncertainty principle. Actually, the latter would imply a factor $\hbar$ at the right-hand side of (5); however, in a one-dimensional space the factor $2\pi$ (half the solid angle of a sphere) is correctly replaced in the proposed model by the factor 1 (half the measure of the unit 1-sphere). The assumption (3) follows from (4) and (5).

Note that several authors have proposed a fundamental discretization of spacetime based on Compton wavelength, including seminal work [32, 33] and, more recently, [34].

## 3 Particle emissions

In this model, particles are generally created (or "emitted") one-by-one at some source nodes ("source") $x_0$, which can be fixed or randomly determined according to a probability mass function (pmf) that represents real scenarios. We assume for simplicity that emissions instants are sufficiently spaced so that at any time there is only one particle traveling. Two pieces of information are attributed to a particle before its emission ("source preparation"): a "source momentum" and a "source phase".

The former, $v_0 \in \mathbb{Q}$, is a uniformly-distributed rational-valued random variable ($v_0 \in \mathbb{Q}$ and $v_0 \approx U[-1,1]$) that is determined at the preparation and does not change during the particle's evolution. This source momentum plays a key role in the proposed model in introducing an intrinsic randomness into the particle's evolution.

The source phase, $\epsilon \in \mathbb{Q}$, is a property of the source node. The particular function $\epsilon(x_0)$ is set in such a way to represent real scenarios. In most cases, it shall give rise to a drift momentum that is summed to $v_0$.

## 4 Microscopic motion

In this section the general characteristics of the random variable $v$ introduced in (2) are described. Since $v$ can take only three values at each time step, its probability distribution is completely determined by two values, its expected value and its variance. In the rest of the paper, expected values will be denoted in bold.

Define the "momentum propensity" as $\boldsymbol{v} := E[v]$. The model further assumes that

$$E[v^2] := \frac{1+\boldsymbol{v}^2}{2} := \boldsymbol{e}, \tag{6}$$

or, in other terms, $\text{Var}[v] = (1 - \boldsymbol{v}^2)/2$. Both $\boldsymbol{v}$ and $\boldsymbol{e}$ are not integers but rational numbers (this point will be clarified later). It should be noticed that, since $\boldsymbol{v} \in [-1,1]$, also $\boldsymbol{e} \in [-1,1]$. The symbol $\boldsymbol{e}$ recalls the fact that this quantity can be regarded as the average value of instantaneous particle's energy and will be denoted as "energy propensity".

Consequently to (6), the probability distribution of $v$ is determined as

$$\Pr(v = 1) = \frac{\boldsymbol{e} + \boldsymbol{v}}{2}, \quad \Pr(v = 0) = 1 - \boldsymbol{e}, \quad \Pr(v = -1) = \frac{\boldsymbol{e} - \boldsymbol{v}}{2}. \tag{7}$$

The quantity $\boldsymbol{v}$ is itself a stochastic variable, resulting from two different mechanisms: (i) imprint during the particle's "preparation" at the source before its emission, and (ii) iteration-by-iteration evolution according to two types of forces, namely, "quantum forces" and "external forces". In summary,

$$\boldsymbol{v}[n] := v_Q[n] + v_F[n], \quad \boldsymbol{v}[n_0] = v_0, \tag{8}$$

where $v_Q$ is the contribution due to quantum forces (it amounts to $v_0$ when these forces are absent, see Sect. 4.2) and $v_F$ is the contribution due to external forces (see Sect. 4.1). It should be noticed that, according to (8), to the fact that $v_0 \in \mathbb{Q}$, and the further rules below, $\boldsymbol{v}$ is a random rational-valued variable as anticipated.

In the absence of either quantum or external forces, the proposed model is summarized as

$$M_0 := \begin{cases} t[n+1] = t[n] + 1, & t[n_0] = 0, \\ x[n+1] = x[n] + v[n], & x[n_0] = x_0, \\ \Pr(v = 0, \pm 1) = (7), \\ \boldsymbol{v}[n] = v_0, \\ v_0 = U[-1; 1]. \end{cases} \tag{9}$$

In this case, the particle just keeps its source momentum and accordingly its evolution can be described by the *average position* $\boldsymbol{x}[n] = x_0 + v_0 t[n]$.

## 4.1 External forces

External forces are described by interactions with the lattice, where each node can be occupied by momentum-mediating entities that will be called "bosons" in analogy with physical force-mediating particles. Depending on their origin, these bosons have an intrinsic momentum propensity $\boldsymbol{v_f}$. The probability of finding such a boson at a certain node, $P_f(x,t)$, depends on the rate at which such bosons are emitted by their source and the distance from the source (the time dependency is because the bosons' source can be variable).

This fundamental mechanism is equivalent to, and for computational easiness replaced by, the following one: bosons are always available at each node where $P_f \neq 0$ and have an intrinsic momentum propensity $f(x,t) := \boldsymbol{v_f} P_f(x,t)$. When a particle visits the lattice node, it captures the "resident" boson and incorporates its momentum. A new boson is then recreated at the node.

The contribution to the particle's momentum propensity due to external forces is thus given by the sum of the momenta of all external bosons captured,

$$v_F[n] = \sum_{n'=n_0+1}^{n} f(x[n'], n').  \qquad (10)$$

It should be noticed that equation (10) is analogous to classical Newton's law in lattice units.

Under the sole action of external forces, the proposed model is summarized as

$$M_1 := \begin{cases} t[n+1] = t[n] + 1, & t[n_0] = 0, \\ x[n+1] = x[n] + v[n], & x[n_0] = x_0, \\ \Pr(v = 0, \pm 1) = (7), & \\ v[n+1] = v[n] + f(x[n], n), & v[n_0] = v_0, \\ v_0 = U[-1; 1]. & \end{cases} \qquad (11)$$

In such scenarios, the momentum propensity varies and accordingly the average position is $x[n] = x_0 + \sum_{n'=n_0}^{n-1} v[n']$.

## 4.2 Quantum forces

The contribution to the momentum propensity due to quantum forces is given by

$$v_Q[n] = v_0 - \sum_i \sum_{j \neq i} v_Q^{(ij)}[n], \qquad (12)$$

where each term in the summation at the right-hand side of (12) results from an exchange of information between the particle and the lattice. In fact, both the particle and the lattice nodes carry and store some integer-valued "counters" that can be updated as iterations proceed.

### 4.2.1 Counter dynamics

The counters carried on by the particle are its lifetime, $t[n]$, a spatial counter $\ell[n]$ denoted as "span", as well its phase $\varepsilon[n]$. The counters stored at each lattice node $\xi$ are the "traces" $\lambda_{\xi\tau}[n]$ and $\varepsilon_{\xi\tau}[n]$, i.e., the memory of the span and phase carried by the last particle that has visited the node with lifetime $\tau$.

The particle span $\ell$ is generally updated at each iteration by summing up the value of the instantaneous momentum. A *sign inversion* occurs when the particle experiences an external force. The particle phase and both lattice traces generally remain constant. However, *when the trace of the lattice node visited is different from the span* ("Reset Condition"), the two counters are interchanged. Similarly, the phase counters are interchanged under the Reset Condition.

In other terms, the dynamics of the spatial counters are given by

$$\ell[n+1] = \begin{cases} \lambda_{x[n]t[n]}[n], & \text{if } C_{\xi\tau} \\ \ell'[n], & \text{else if } f(x[n], n) = 0 \\ -\ell'[n], & \text{otherwise if } f(x[n], n) \neq 0 \end{cases}, \qquad (13)$$

$$\lambda_{\xi\tau}[n+1] = \begin{cases} \ell'[n], & \text{if } C_{\xi\tau} \\ \lambda_{\xi\tau}[n], & \text{otherwise} \end{cases}. \qquad (14)$$

where $\ell[n_0] = 0$, $\ell'[n] := \ell[n] + v[n]$, and where the reset condition is defined as $C_{\xi\tau} := (x[n] = \xi) \wedge (t[n] = \tau) \wedge (\ell[n] \neq \lambda_{\xi\tau}[n-1])$.

According to these rules, it should be clear that the trace found by a particle can be different from its span because the last particle that visited the node with the same lifetime either had been emitted from a different source $x_0$ or had captured a different number of external bosons. In any case, it should be noticed that $\ell[n] \in \mathbb{Z}$. Consequently, also $\lambda_{\xi\tau}[n] \in \mathbb{Z}$.

### 4.2.2  Boson creation

The Reset Condition also creates a new momentum-carrying "lattice boson" (LB). This boson is labelled with the particular pair of integers $\ell[n], \lambda_{\xi\tau}[n]$ or, equivalently, with the pair $ij$, where $i := \xi - \ell[n]$ and $j := \xi - \lambda_{\xi\tau}[n]$. Clearly, $i \in \mathbb{Z}$ and $j \in \mathbb{Z}$ are images of the respective sources of the current particle and of the last particle that has visited the node $\xi$ with the same lifetime.

This LB replaces the previously resident boson of the same type, if there was one. The latter, before being replaced, is transferred to the particle (hence it becomes a "particle boson", PB).

The new LB is created with a momentum (LBM)

$$\bar{\omega}_{\xi\tau}^{(ij)}[n] = \left\{ \delta^{(ij)} v_Q[n-1] - \varepsilon^{(ij)} \right\}, \tag{15}$$

that is, a fraction of the quantum momentum of the visiting particle and its phase, with the "path difference" defined as

$$\delta^{(ij)} := |i - j| = \left| \ell[n] - \lambda_{\xi\tau}[n-1] \right|. \tag{16}$$

and the "phase difference", resulting from a different preparation at the two sources, is

$$\varepsilon^{(ij)} := \varepsilon[n] - \varepsilon_{\xi\tau}[n-1]. \tag{17}$$

The function $\{\cdot\}$ stands here for a shifted "modulo 2" operation, that is, $\{x\} = -1 + \mathrm{mod}(x+1, 2)$.

Conversely, the new particle boson momentum (PBM) equals the old LBM divided by the path difference,

$$v_Q^{(ij)}[n] = \frac{\omega_{x[n]t[n]}^{(ij)}[n-1]}{\delta^{(ij)}}, \tag{18}$$

and contributes to the right-hand side of (11).

### 4.2.3  Boson dynamics

When the Reset Condition does not occur, particle and lattice bosons are not replaced. Their momenta, however, decay with the respective boson's lifetimes: at a new iteration, the momentum is only a fraction of the previous value. Particle boson momentum decays as the inverse of its lifetime. Lattice boson momentum decays as the inverse square of its lifetime.

The whole mechanism can be formalized as follows. The PBM dynamics is given by

$$v_Q^{(ij)}[n] = \begin{cases} v_Q^{(ij)}[n-1] \cdot \left(1 - \dfrac{1}{2k^{(ij)}[n]}\right), & \text{if } k^{(ij)}[n] > 0 \\ \dfrac{\omega_{x[n]t[n]}^{(ij)}[n]}{\delta^{(ij)}}, & \text{otherwise} \end{cases}, \tag{19}$$

$$k^{(ij)}[n] = \begin{cases} 0, & \text{if } C_{\xi\tau}^{(ij)} \\ k^{(ij)}[n-1] + 1, & \text{else} \end{cases}, \tag{20}$$

where $k^{(ij)}$ is the lifetime of the $ij$-boson, the reset condition (when the boson transfer takes place) is

$$C_{\xi\tau}^{(ij)} := (x[n] = \xi) \wedge (t[n] = \tau) \wedge (\ell[n] = \xi - i) \wedge (\lambda_{\xi\tau}[n-1] = \xi - j). \tag{21}$$

The LBM dynamics is given by the rules

$$\omega_{\xi\tau}^{(ij)}[n] = \begin{cases} \omega_{\xi\tau}^{(ij)}[n-1] \cdot \left(1 - \left(\dfrac{\bar{\omega}_{\xi\tau}^{(ij)}[n]}{\kappa_{\xi\tau}^{(ij)}[n]}\right)^2\right), & \kappa_{\xi\tau}^{(ij)}[n] > 0 \\ \bar{\omega}_{\xi\tau}^{(ij)}[n], & \text{otherwise} \end{cases}, \tag{22}$$

$$\bar{\omega}_{\xi\tau}^{(ij)}[n] = \begin{cases} \bar{\omega}_{\xi\tau}^{(ij)}[n-1], & \kappa_{\xi\tau}^{(ij)}[n] > 0 \\ \delta^{(ij)} v_Q[n] - \varepsilon^{(ij)}, & \text{otherwise} \end{cases}, \tag{23}$$

$$\kappa_{\xi\tau}^{(ij)}[n] = \begin{cases} 0, & \text{if } C_{\xi\tau}^{(ij)} \\ \kappa_{\xi\tau}^{(ij)}[n-1] + 1, & \text{otherwise} \end{cases}, \tag{24}$$

where $\kappa_{\xi\tau}^{(ij)}$ is the lifetime of the lattice $ij$-boson, $\bar{\omega}_{\xi\tau}^{(ij)}$ is its initial momentum (the boson has a memory of it). It should be noticed that the rules above preserve the fact that $v_Q \in \mathbb{Q}$, $\bar{\omega}_{\xi\tau}^{(ij)} \in \mathbb{Q}$, and $v_Q^{(ij)} \in \mathbb{Q}$.

The complete set of equations of the proposed model is summarized as follows:

$$M := \begin{cases} t[n] = t[n-1] + 1, & t[n_0] = 0, \\ x[n+1] = x[n] + v[n], & x[n_0] = x_0, \\ \Pr(v = 0, \pm 1) = (7), \\ v[n] := v_Q[n] + v_F[n], & v[n_0] = v_0, \\ v_0 = U[-1; 1], \\ v_F[n] = \displaystyle\sum_{n'=n_0+1}^{n} f(x[n'], n'), \\ v_Q[n] = v_0 - \displaystyle\sum_i \sum_{j \neq i} v_Q^{(ij)}[n], \\ v_Q^{(ij)}[n] = (19) - (24). \end{cases} \tag{25}$$

Table 1

| | $n$ | 1 | 2 | 3 | 4 | 5 | 6 |
|---|---|---|---|---|---|---|---|
| Particle motion, $v_0 = 0.3$, $x_0 = 2$ | $x$ | 3 | 3 | 3 | 2 | 1 | 2 |
| | $\ell'$ | 1 | 1 | 1 | 2 | 1 | 2 |
| | $\ell$ | 1 | 1 | 3 | 2 | 1 | 0 |
| | RC | - | - | 1 | 0 | 0 | 1 |
| | $ij$ | - | - | 20 | - | - | 02 |
| Lattice trace | $\lambda_{33}$ | 3 | 3 | 1 | 1 | 1 | 1 |
| | $\lambda_{42}$ | 2 | 2 | 2 | 2 | 2 | 2 |
| | $\lambda_{51}$ | 1 | 1 | 1 | 1 | 1 | 1 |
| | $\lambda_{62}$ | 0 | 0 | 0 | 0 | 0 | 2 |
| L-boson momentum (*) | $\omega_{33}$ | ... | 0.2 | 0.6 | 0.6 | ... | ... |
| | $\omega_{42}$ | ... | ... | 0.1 | 0.1 | 0.1 | ... |
| | $\omega_{51}$ | ... | ... | ... | 0 | 0 | 0 |
| | $\omega_{62}$ | ... | ... | ... | ... | -0.1 | 0.525 |
| P-boson momentum | $v_Q^{(20)}$ | - | - | 0.1 | 0.05 | 0.0375 | 0.0312 |
| | $k^{(20)}$ | - | - | 0 | 1 | 2 | 3 |
| | $v_Q^{(02)}$ | - | - | - | - | - | -0.05 |
| | $k^{(02)}$ | - | - | - | - | - | 0 |
| | $v_Q$ | 0.3 | 0.3 | 0.2 | 0.25 | 0.2625 | 0.3188 |

The proposed mechanism is illustrated by the example in Table 1. A particle is emitted at $n = 0$ from a source at $x = 2$, while another source at $x = 0$ was possible. The particle trajectory is followed for 6 iterations. The span $\ell'$ is computed and compared with the trace at the lattice site visited by the particle at each iteration. If they differ, the RC is set to one and the two counters $\ell$ and $\lambda_{x[n]t[n]}$ exchanged (red numbers). Otherwise, the RC is false (green numbers). In the RC case, the boson label $ij$ is defined; a LB and a PB are created or updated, with respective momenta initialized according to the rules above (red numbers $\omega_{x[n]t[n]}$, $v_Q^{(ij)}$). Otherwise, the existing LBM and PBM decay (green numbers). Finally, PBM contribute to the particle's quantum momentum.

## 5 Probability densities

In QM, probability densities of observables are evaluated from complex wavefunctions that are solutions of Schrödinger's equation. In turn, in the proposed model the probability mass functions or densities are calculated directly from the motion rules.

Even without quantum or external interactions, the fact that the source momentum is a random variable implies that $v$ and thus $x$ are random variables, too. The probability mass function $\rho(x; t)$ is evaluated from an ensemble of similarly-prepared particles.

The probability mass function of the source momentum is

$$\rho(v_0) = \frac{1}{2}. \tag{26}$$

Additionally, the source location is treated as a random variable, too. In general, there are $N_s$ possible sources, located at nodes $x_0^{(k)}$, each of which has a probability $P_0^{(k)}$. In other terms,

$$x_0 = \left\{ x_0^{(k)} \right\}, \qquad k = 1, \dots, N_s, \qquad \rho(x_0) = P_0^{(k)} \cdot \delta\left(x - x_0^{(k)}\right), \tag{27}$$

where $\delta(\cdot)$ is a Kronecker delta function. Four special cases are considered for the sake of presentation: (i) no forces, (ii) only quantum forces, (iii) only homogeneous external forces from a quadratic potential, and (iv) quantum and homogeneous external forces from a quadratic potential.

It turns out that for all these cases the expected value of the position is a monotonic function of the quantum momentum and the latter of the source momentum. However, the quantum momentum is not an explicit function of the source location. The chain rule

$$\rho(\pmb{x};t) = \rho(v_0) \left|\frac{dv_0}{d\pmb{x}}\right| = \rho(v_0) \left|\frac{dv_0}{dv_Q}\right| \left|\frac{dv_Q}{d\pmb{x}}\right| \tag{28}$$

is thus applied.

## 5.1 No forces

The only scenario without forces acting on the particle is when there is a single source possible $x_0$ and no external forces. In this scenario, each lattice node $\xi$ always receives particles carrying a span equal to $\xi - x_0$, so that no bosons are created.

For illustration purposes, the probability mass function of the position can be explicitly evaluated in this case. For a given $\pmb{v} = v_0$, the pmf of $x$ at a given $t$ is

$$w(x, \pmb{v}; t, x_0) = \frac{\binom{2t}{t + x - x_0}}{2^{2t}} (1 + \pmb{v})^{t+x-x_0} (1 - \pmb{v})^{t-x+x_0} \tag{29}$$

and is well approximated by a Gaussian function $\frac{1}{\sqrt{2\pi\mathcal{D}t}} \exp\left(-\frac{(x-x_0-\pmb{v}t)^2}{2\mathcal{D}t}\right)$, where $\mathcal{D} := (1 - \pmb{v}^2)/2$. By integrating over values of $v_0$, we obtain

$$\rho(x;t) = \int_{-1}^{1} \rho(v_0) w(x, v_0; t, x_0) dv_0 = \frac{1}{2t + 1}, \tag{30}$$

that is, a constant pmf in the reachable interval, $x = U[x_0 - t, x_0 + t]$.

This result can be approximated by using (28) and observing that $\pmb{x} = x_0 + v_0 t$ in this case. Therefore

$$\rho(\pmb{x}; t) = \frac{1}{2t}. \tag{31}$$

It should be noticed that $\rho(x; t) \approx \rho(\pmb{x}; t)$ for large times. This approximation will be used in the following scenarios, where it is generally not possible to explicitly evaluate $\rho(x; t)$. However, it should be noticed that, while $x \in \mathbb{Z}$, its expected value will be approximated by real numbers ($\pmb{x} \in \mathbb{R}$) in the following, at least for large times.

For the no-forces scenario it is also possible to explicitly compute the pmf of another random variable defined for each lattice node as

$$\sigma_{\xi\tau}[n] := \begin{cases} s[n], & \text{if } (x[n] = \xi) \wedge (t[n] = \tau) \\ \sigma_{\xi\tau}[n - 1], & \text{otherwise} \end{cases}, \tag{32}$$

The particle random variable $s$ is in turn defined recursively as

$$s[n + 1] = s[n] + |v[n]|, \quad s[n_0] = 0, \tag{33}$$

and can be regarded as the accumulated energy of the particle, in agreement with the fact that the expected value of $|v|$ is the energy propensity $\pmb{e}$ defined above, and will be referred to here as the

particle's "action", at least for this special case (a term due to external bosons is actually to be introduced when external forces are acting). The variable $\sigma_{\xi\tau}$ is the particle action "seen" by the node when particles visit it. It should be further noticed that both $s$ and $\sigma_{\xi\tau} \in \mathbb{N}$.

The pmf of the action seen at a node can be explicitly evaluated for this simple scenario as

$$\rho_v(\sigma; x, t) = \frac{2^{t-s} \binom{\frac{\sigma + x - x_0}{2}}{\frac{t}{t - \sigma}} \binom{t - \frac{\sigma + x - x_0}{2}}{t - \sigma}}{\binom{2t}{t + x - x_0}}, \quad (34)$$

where $\sigma \in \left\{|x - x_0|, |x - x_0| + 2, \ldots, |x - x_0| + 2\left\lfloor\frac{t - |x - x_0|}{2}\right\rfloor\right\}$ (subscripts $\xi\tau$ have been omitted here for the sake of clarity). Since the right-hand side of (34) does not depend on $v_0$, $\rho(\sigma; x, t) = \rho_v(\sigma; x, t)$ holds as well.

The expected value of the action seen at a node is found with some algebraic manipulation to be

$$\sigma_{xt} = \frac{(x - x_0)^2 + t^2 - t}{2t - 1}, \quad (35)$$

which, for large times, is remarkably similar to the classical free particle action $S(x, t) = \frac{(x - x_0)^2}{2t}$ plus the term $t/2$.

In conclusion, the proposed model approximates for large times the probability density and the action of a free particle emitted from a single source, albeit only using integer and rational quantities.

## 5.2  Quantum forces only

When source location can take multiple values, quantum forces occur. In fact, a lattice node $\xi$ can receive particles carrying a span that takes either of the values $\xi - x_0^{(k)}$, so that bosons are created. We shall consider the generic $ij$-bosons.

### 5.2.1  Lattice training

*Lattice training* is the process during which the LBM $\omega_{\xi\tau}^{(ij)}$ tends to its expected value. Generally, $\omega_{\xi\tau}^{(ij)}$ depends on the boson's lifetime according to the decay rule (22). By repeatedly applying such a rule for $\kappa$ iterations, we obtain

$$\omega_{\xi\tau}^{(ij)}(\kappa) = \overline{\omega}_{\xi\tau}^{(ij)} \prod_{\kappa'=1}^{\kappa} \left(1 - \left(\frac{\overline{\omega}_{\xi\tau}^{(ij)}}{\kappa'}\right)^2\right). \quad (36)$$

We can safely assume that the site is visited only rarely by a particle, given the generally huge number of sites and assuming that a sufficiently long time passes between two successive emissions (we assumed that at one time there is only one particle traveling). Under this assumption, the expected value of the LBM $\omega_{\xi\tau}^{(ij)}$ tends to coincide with its steady-state value, obtained by letting $\kappa$ tend to infinity. Using the known formula for the sine expansion, $\sin \pi z = \pi z \prod_{n=1}^{\infty}(1 - z^2/n^2)$, it turns out that

$$\boldsymbol{\omega}_{\xi\tau}^{(ij)} = \frac{\sin\left(\pi\overline{\omega}_{\xi\tau}^{(ij)}\right)}{\pi}. \tag{37}$$

It should be remarked that a trigonometric functionality emerges quite naturally from the integer-valued model proposed.

As for the initial LBM $\overline{\omega}_{\xi\tau}^{(ij)}$, it is a stochastic variables that can change only at time $\tau$ of each emission according to rule (15). It is clear that after a sufficiently large number of iterations the sample mean of $\overline{\omega}_{\xi\tau}^{(ij)}$ tends to the expected value

$$\overline{\boldsymbol{\omega}}_{\xi\tau}^{(ij)} = \delta^{(ij)}\frac{\left(\xi - x_0^{(i)}\right) + \left(\xi - x_0^{(j)}\right)}{2\tau} - \epsilon^{(ij)} = \delta^{(ij)}\frac{\xi - \frac{x_0^{(i)} + x_0^{(j)}}{2}}{\tau} - \epsilon^{(ij)} = \overline{\boldsymbol{\omega}}_{\xi\tau}^{(ji)}, \tag{38}$$

and consequently $\boldsymbol{\omega}_{\xi\tau}^{(ij)}$ to a value

$$\boldsymbol{\omega}_{\xi\tau}^{(ij)} = \frac{\sin\left(\pi\overline{\boldsymbol{\omega}}_{\xi\tau}^{(ij)}\right)}{\pi}. \tag{39}$$

This process is illustrated in Figure 1-Figure 2.

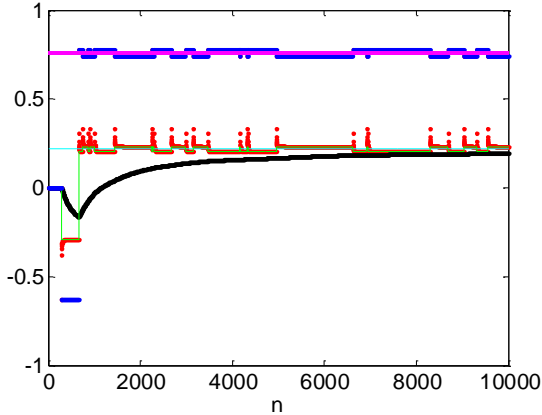

Figure 1: Outcome of one simulation ($x_0 = \{100 \pm 1\}$, $P_0 = \{0.5, 0.5\}$) in terms of $\overline{\omega}_{xt}^{(12)}$ (blue), $\omega_{xt}^{(12)}$ (red), its running average (black), $\omega_{xt}^{(12)}$ from (37) (green), $\overline{\omega}_{xt}^{(12)}$ from ((38) (magenta), and $\omega_{xt}^{(12)}$ from (39) (cyan), for a node ($x = 38$, $t = 100$) as a function of the number of iterations.

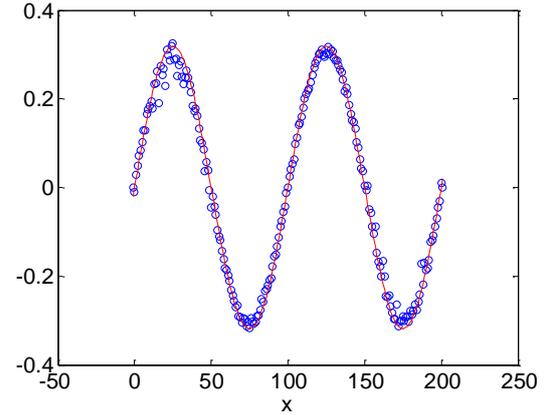

Figure 2: Outcome of the same simulation of Figure 1 (after $N_p = 50000$): $\boldsymbol{\omega}_{xt}^{(12)}$ (blue), theoretical values (red) for each lattice node.

### 5.2.2 Particle training

*Particle training* is the process during which the expected value of the PBM $v_Q^{(ij)}$ stabilizes. Generally, $v_Q^{(ij)}$ varies with the boson's lifetime, according to rule (19). The probability that such a boson has lifetime $k$ is equal to the probability that in $k$ iterations a new boson is created only once. The probability that a new boson is created equals that of the joint event $P^{(ij)} := P_0^{(i)} P_0^{(j)} = P^{(ji)}$. Therefore, $\Pr\left(k^{(ij)} = k\right) = P^{(ij)}\left(1 - P^{(ij)}\right)^k$. The expected value of $v_Q^{(ij)}$ is thus evaluated as

$$v_Q^{(ij)} = P^{(ij)} \sum_{k=0}^{\infty} v_Q^{(ij)}(k) \cdot \left(1 - P^{(ij)}\right)^k, \quad (40)$$

where $v_Q^{(ij)}(k)$ denotes now the PBM with lifetime $k$. By repeatedly applying rule (19), we arrive at

$$v_Q^{(ij)}(k) = v_Q^{(ij)}(0) \cdot \prod_{k'=1}^{k} \frac{2k'-1}{2k'}. \quad (41)$$

The product in (41) is evaluated as $\frac{(2k)!}{(k!)^2 4^k}$ that, for the properties of Gamma function, is formally equivalent to $(-1)^k \binom{-1/2}{k}$. Therefore, (40) is manipulated as

$$v_Q^{(ij)} = P^{(ij)} \sum_{k=0}^{\infty} \left(1 - P^{(ij)}\right)^k \cdot v_Q^{(ij)}(0) \cdot (-1)^k \binom{-1/2}{k} =$$
$$= v_Q^{(ij)}(0) P^{(ij)} \left(1 - \left(1 - P^{(ij)}\right)\right)^{-1/2} = v_Q^{(ij)}(0)\sqrt{P^{(ij)}}, \quad (42)$$

having used the binomial series expansion $(1-z)^\alpha = \sum_{k=0}^{\infty} \binom{\alpha}{k}(-z)^k$.

The next step consists of replacing $v_Q^{(ij)}(0)$ with $\omega_{xt}^{(ij)}/\delta^{(ij)}$, according to rule (19) and with the change of subscripts $\xi \to x$. Using (39), we find

$$v_Q^{(ij)} = \sqrt{P^{(ij)}} \frac{\sin\left(\pi \bar{\omega}_{xt}^{(ij)}\right)}{\pi \delta^{(ij)}}. \quad (43)$$

This process is illustrated in Figure 3-Figure 4, where one random variable $v_Q^{(ij)}$ is shown versus the number of iterations for one emission, together with its running average and the quantity $v_Q^{ij}$ expected from (43). The figure clearly shows that the running average tends after a sufficiently long time to the expected value. Consequently, also the "average momentum" calculated as $(x - x_0)/t$ tends to $v_Q$ given by (44).

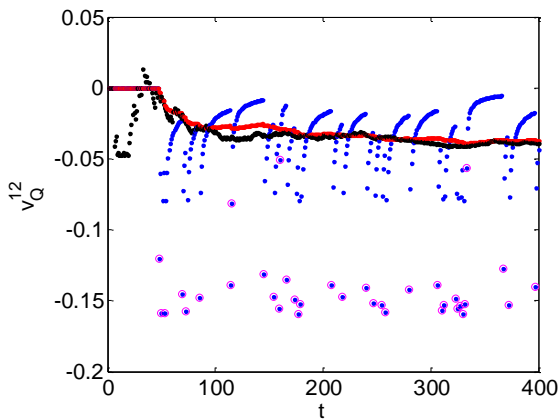
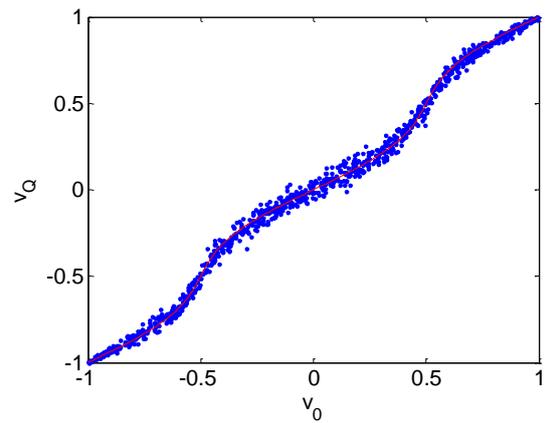

Figure 3: Outcome of one simulation ($N_t = 1000$, $v_0 = .7845$, $x_0 = \{\pm 1\}$, $P_0 = \{0.1, 0.9\}$): $v_Q^{(12)}(0)$ (magenta), $v_Q^{(12)}$ (blue), its running average (red), $v_Q^{(12)}$ from (43) (black), as a function of particle's

Figure 4: Outcome of the same simulation of Figure 3 (for $N_p = 50000$ emissions): $v_Q$ (blue), theoretical values (red) as a function of $v_0$.

lifetime.

### 5.2.3 Position and momentum pdf

The expected value of the particle's (total) quantum momentum is eventually found from (28) as

$$\boldsymbol{v_Q} = v_0 - \sum_i \sum_{j \neq i} \sqrt{P_0^{(i)} P_0^{(j)}} \frac{\sin\left(\pi \delta^{(ij)} \frac{\boldsymbol{x} - \frac{x_0^{(i)} + x_0^{(j)}}{2}}{t} - \pi \epsilon^{(ij)}\right)}{\pi \delta^{(ij)}}. \tag{44}$$

The pmf of the position cannot be explicitly evaluated in this scenario. However, the probability density function of its expected value can be evaluated, using (44) and observing that $\boldsymbol{x} = x_0 + \boldsymbol{v_Q} t$ holds in this case, as

$$\rho(\boldsymbol{x}; t) = \frac{1}{2} \frac{dv_0}{d\boldsymbol{x}} = \frac{1 + \sum_i \sum_{j \neq i} \sqrt{P_0^{(i)} P_0^{(j)}} \cos\left(\pi \delta^{(ij)} \frac{\boldsymbol{x} - \frac{x_0^{(i)} + x_0^{(j)}}{2}}{t} - \pi \epsilon^{(ij)}\right)}{2t}. \tag{45}$$

A further analysis of (45) remarkably allows to retrieve Schrödinger equation and Born rule. First replace $\delta^{(ij)}$ with $\left| x_0^{(i)} - x_0^{(j)} \right|$ and then recognize that the cosine argument divided by $\pi$ is equal to

$$\frac{2\boldsymbol{x}\left(x_0^{(i)} - x_0^{(j)}\right) - \left(\left(x_0^{(i)}\right)^2 - \left(x_0^{(j)}\right)^2\right)}{2t} - \epsilon^{(ij)} = -\Delta S^{(ij)}(\boldsymbol{x}, t). \tag{46}$$

where the function $S^{(k)}(x, t)$ is the classical free-particle action with respect to the $k$-th source,

$$S^{(k)}(x, t) := \frac{\left(x - x_0^{(k)}\right)^2}{2t} + \epsilon^{(k)}. \tag{47}$$

The right-hand side of (45) can be then equivalently obtained as the square modulus of a complex number, $\rho(\boldsymbol{x}; t) = |\psi(x,t)|^2$, provided that $\psi$ is defined as

$$\psi(x,t) := \sum_k \psi^{(k)}(x,t), \qquad \psi^{(k)}(x,t) := \sqrt{\frac{P_0^{(k)}}{2t}} \exp(\iota \pi S^{(k)}(x,t)). \tag{48}$$

It is easy to recognize $\psi(x, t)$ as the probability amplitude of the free particle having many possible sources, each of which has a probability amplitude $\psi^{(k)}(x, t)$ satisfying Schrödinger equation. With this observation, the equivalence between the proposed model and quantum mechanics is demonstrated for any free-particle scenario.

## 5.3 Homogeneous external forces only (quadratic potentials)

This section treats the scenario where particles are emitted from a single source as in Sect. 5.1; however, particles are now subject to external forces. The analysis is limited to quadratic potentials such that the external boson momentum is

$$f(x, t) = \alpha(t) x + \beta(t). \tag{49}$$

It should be noticed that (49) is applied to every lattice node $x$. Consequently, each node visited by the particle transmits an external boson and, according to rule (14),

$$\ell[n] = (-1)^{\text{parity}(t[n])} \cdot \frac{(x[n] - x_0)}{t[n]}. \tag{50}$$

According to that, the span carried by the particle at a given node only depends on its lifetime. Therefore there are no possible differences between the span and the trace found that might be induced by external forces. Consequently, quantum forces are always null, $v_Q[n] \equiv v_0$ and $v[n] = v_0 + v_F[n]$.

Computing the pdf of $x$ requires the particularization of the function $f(x,t)$ that describes the EBM. Generally speaking, for quadratic potentials it is always true that

$$x(t) = A(t)x_0 + B(t)v_0 + C(t), \tag{51}$$

where $A(t)$, $B(t)$, and $C(t)$ are functions of lifetime whose form depends on the coefficients $\alpha$ and $\beta$ of (49), such as

$$\begin{cases} A(0) = 1, & \dot{A}(0) = 0, & \ddot{A}(t) = \alpha(t)A(t), \\ B(0) = 0, & \dot{B}(0) = 1, & \ddot{B}(t) = \alpha(t)B(t), \\ C(0) = 0, & \dot{C}(0) = 0, & \ddot{C}(t) = \alpha(t)C(t) + \beta(t), \end{cases} \tag{52}$$

and

$$\dot{A}B = A\dot{B} - 1. \tag{53}$$

Some examples of external forces will be presented in Sect. 9. In general terms, the pdf of the expected value of the particle position can be evaluated as

$$\rho(x;t) = \frac{1}{2}\left|\frac{dv_0}{dx}\right| = \frac{1}{2|B(t)|}. \tag{54}$$

The accessible domain of the particle position is limited by the trajectories obtained by setting $v_0 = \pm 1$ in (51). Consequently, it can be verified that

$$\int_{A(t)x_0+B(t)+C(t)}^{A(t)x_0-B(t)+C(t)} \rho(x;t)dx = 1. \tag{55}$$

Note that (54) coincides with the Van Vleck determinant for the average motion, $\left|\frac{\partial^2 S}{\partial x_0 \partial x}\right|$ since for quadratic potentials $v_0 = -\frac{\partial S}{\partial x_0}$.

### 5.4 Quantum and homogeneous external forces (quadratic potentials)

In this scenario the particle encounters both quantum and external bosons. Consequently, the role of $v_0$ in (51) is now played by the quantum momentum and thus that equation is replaced by

$$x(t) = A(t)x_0 + B(t)v_Q + C(t), \tag{56}$$

where $x_0$ is, as in Sect. 5.2, a random variable.

To evaluate the pdf of the expected position, first apply rule (23) that yields

$$\overline{\omega}_{xt}^{(ij)} = \delta^{(ij)} \frac{x - A(t)\frac{x_0^{(i)} + x_0^{(j)}}{2} - C(t)}{B(t)} - \epsilon^{(ij)}, \tag{57}$$

Then, using the same reasoning as in Sect. 5.2, the expected value of the quantum momentum is evaluated as

$$\boldsymbol{v_Q} = v_0 - \sum_i \sum_{j \neq i} \sqrt{P_0^{(i)} P_0^{(j)}} \frac{\sin\left(\pi \delta^{(ij)} \frac{\boldsymbol{x} - A(t) \frac{x_0^{(i)} + x_0^{(j)}}{2} - C(t)}{B(t)} - \pi \epsilon^{(ij)}\right)}{\pi \delta^{(ij)}}, \quad (58)$$

that is, as a generalization of (44). The final step is to generalize (45) and find

$$\rho(\boldsymbol{x}; t) = \frac{1 + \sum_i \sum_{j \neq i} \sqrt{P_0^{(i)} P_0^{(j)}} \cos\left(\pi \delta^{(ij)} \frac{\boldsymbol{x} - A(t) \frac{x_0^{(i)} + x_0^{(j)}}{2} - C(t)}{B(t)} - \pi \epsilon^{(ij)}\right)}{2|B(t)|}. \quad (59)$$

Similarly to Sect. 5.2, it should be noticed that the cosine argument divided by $\pi$ can be expressed as

$$\frac{2\left(x_0^{(i)} - x_0^{(j)}\right) x - A(t)\left(\left(x_0^{(i)}\right)^2 - \left(x_0^{(j)}\right)^2\right) - 2C(t)\left(x_0^{(i)} - x_0^{(j)}\right)}{2B(t)} - \epsilon^{(ij)} = -\Delta S^{(ij)}(\boldsymbol{x}, t), \quad (60)$$

where

$$S^{(k)}(x, t) = \frac{1}{2B(t)}\left[A(t)\left(x_0^{(k)}\right)^2 - 2x x_0^{(k)} + 2C(t) x_0^{(k)} + \dot{B}(t) x^2 \right. \\ \left. + \left(2\dot{C}(t)B(t) - 2\dot{B}(t)C(t)\right)x + 2C^2(t)\dot{B}(t)\right] + \epsilon^{(k)} \quad (61)$$

is the classical action in the presence of the considered potential, as it can be easily verified (in particular, verify that $v_0 = -\frac{\partial S}{\partial x_0}$ and $\frac{\partial^2 S}{\partial x_0 \partial x} = -\frac{1}{B}$).

The right-hand side of (59)(61) can be then equivalently obtained as the square modulus of a complex number, $\rho(\boldsymbol{x}; t) = |\psi(x,t)|^2$, where

$$\psi(x, t) = \sum_k \psi^{(k)}(x, t), \qquad \psi^{(k)}(x, t) := \sqrt{\frac{P_0^{(k)}}{2t}} \exp\left(\iota \pi S_{cl}^{(k)}(x, t)\right). \quad (62)$$

It is easy to recognize in $\psi(x, t)$ the probability amplitude of the particle for many possible sources, each of which has a probability amplitude $\psi^{(k)}(x, t)$. With this observation, the equivalence between the proposed model and quantum mechanics is demonstrated also for the scenario considered.

## 5.5 Quantum and inhomogeneous external forces, including potential barriers

For non-quadratic potentials equation (51) is no longer valid. Moreover, the classical action $S$ is no longer a one-valued function of the position. Therefore it is not possible (at least not in an easy way) to explicitly show the correspondence between the proposed mechanism and QM results, i.e., to derive an equation similar to (62).

The most complex scenario considered here is the case of inhomogeneous external forces, where the EBM, i.e., the function $f(x, t)$, does not concern all the possible lattice nodes $x$. In such cases, the sign of the span $\ell$ depends on the path taken, and more precisely on the number of external bosons encountered. Therefore, even for particles emitted from a single source, different spans can be monitored at a given lattice node. As a consequence, quantum forces arise.

Infinite potentials are limit cases of the inhomogeneous external force scenario. They can be conveniently represented as geometric constraints applied at certain nodes (denoted below as the set $\mathcal{X}_B$), where the sign of the quantum momentum (including the source momentum and all the particle bosons) and of the span are instantaneously inverted,

$$f[x \in \mathcal{X}_B, n] = -2v_Q[n-1], \quad . \tag{63}$$
$$\ell[n] = -\ell[n-1] \quad \text{if } x[n] \in \mathcal{X}_B$$

Generally speaking, these problems are equivalent to introducing a certain number of image sources, each with its own probability and phase, then treating the homogeneous external force scenario resulting with such virtual sources. The considerations of Sect. 5.4 then apply.

## 5.6 Stationary States

Stationary states are particular source preparations whose evolution preserves the source probability. A general expression for such probability function $P_{ss}$ is

$$P_{ss}(\boldsymbol{x}) = \frac{1 + \sum_i \sum_{j \neq i} \sqrt{P_{ss}\left(x_0^{(i)}\right) P_{ss}\left(x_0^{(j)}\right)} \cos\left(\pi \delta^{(ij)} \frac{\boldsymbol{x} - A(t)\frac{x_0^{(i)} + x_0^{(j)}}{2} - C(t)}{B(t)} - \pi \epsilon^{(ij)}\right)}{2B(t)}. \tag{64}$$

In principle this formula makes possible to find the pdf of stationary states for a given source set and external forces. However, due to its complexity, it is of little practical usefulness, even when replacing the double sum with a double integral.

As a relatively simple example, it is possible to show that for a harmonic oscillator the $n=0$ stationary state $P_0(x) = \sqrt{\Omega} e^{-\pi \Omega x^2}$ obeys (64) since

$$\int_{-\infty}^{\infty} \int_{-\infty}^{\infty} \left\{ \sqrt{\Omega} e^{-\frac{\pi \Omega (y^2 + z^2)}{2}} \cos\left(\frac{\pi \Omega}{\sin \Omega t}\left(\frac{y^2 - z^2}{2} \cos \Omega t - x(y-z)\right)\right)\right\} dy\, dz$$
$$= \frac{2}{\sqrt{\Omega}} e^{-\pi \Omega x^2} \sin \Omega t = P_0(x) \cdot 2B(t), \tag{65}$$

with $B(t) = \sin \Omega t / \Omega$.

## 5.7 Momentum probability density

For quadratic potentials, for which $\boldsymbol{x} = A(t)x_0 + B(t)\boldsymbol{v_Q}(t) + C$, the momentum pdf is evaluated (convolution of probability distributions) as

$$\rho_{v_Q}(\boldsymbol{v_Q}; t) = |B(t)| \sum_k P_0^{(k)} \rho\left(A(t) x_0^{(k)} + B(t) \boldsymbol{v_Q} + C; t\right), \tag{66}$$

where $\rho(\boldsymbol{x}; t)$ is given by (59). An approximation that is valid in most cases consists in neglecting the differences $x_0 - \frac{x_0^{(i)} + x_0^{(j)}}{2}$, thus letting $\rho$ be independent of the source. That leads to a steady-state pdf

$$\rho_{v_Q}(\boldsymbol{v_Q}) \approx \frac{1}{2}\left(1 + \sum_i \sum_{j \neq i} \sqrt{P_0^{(i)} P_0^{(j)}} \cos\left(\pi \delta^{(ij)} \boldsymbol{v_Q} - \pi \epsilon^{(ij)}\right)\right). \tag{67}$$

For the $n=0$ harmonic oscillator, this formula can be integrated and yields $\rho_{v_Q}(\boldsymbol{v_Q}) \approx \frac{1}{\sqrt{\Omega}} e^{-\frac{\pi v_Q^2}{\Omega}}$.

Of course, since $v_Q = (x - A(t)x_0 - C(t))/B(t)$ and $\sum_k P_0^{(k)} = 1$, the relationship

$$\rho(x;t) = \frac{1}{B(t)} \sum_k P_0^{(k)} \cdot \rho_{v_Q}\left(\frac{x - A(t)x_0^{(k)} - C(t)}{B(t)}; t\right), \tag{68}$$

also holds, as it is easy to verify. On the other hand, (68) is also obtained by generalizing (30) to the most general case. Equations (67) and (68) thus play the same role in the proposed model than the Fourier's transforms in QM.

## 5.8 Eigenvalues and quantization
The pdf (66) might have definite peak values (corresponding to the quantized values or the "eigenvalues" of QM), which are obtained by setting $\partial \rho_{v_Q}(v_Q)/\partial v_Q = 0$ or, equivalently, $\partial^2 v_0/\partial v_Q^2 = 0$. We shall label these peak values $\hat{v}_Q$. Consequently, peak values of kinetic energy and angular momentum will be labelled $\hat{E}$ and $\hat{J}$, respectively.

## 5.9 Wigner function
The function $W(x, v_Q; t) = \sum_{x_0} P(x_0) \cdot w(x, v_Q; t, x_0) \cdot \rho_{v_Q}(v_Q)$ is a definite-positive Wigner function. In fact, using the Dirac-like properties of the function $w(\cdot)$, it is easily verified that

$$\begin{aligned}
\int W(x, v_Q; t) dv_Q &= \int \sum_{x_0} P(x_0) \cdot w(x, v_Q; t, x_0) \cdot \rho_{v_Q}(v_Q) \, dv_Q \\
&= \sum_{x_0} P(x_0) \int w(x, v_Q; t, x_0) \cdot \rho_{v_Q}(v_Q) dv_Q \\
&= \sum_{x_0} P(x_0) \cdot \rho_{v_Q}\left(\frac{x - A(t)x_0 - C(t)}{B(t)}\right) \cdot \frac{1}{B(t)} = \rho(x; t),
\end{aligned} \tag{69}$$

and, on the other hand, that

$$\begin{aligned}
\int W(x, v_Q; t) dx &= \int \sum_{x_0} P(x_0) \cdot w(x, v_Q; t, x_0) \cdot \rho_{v_Q}(v_Q) \, dx \\
&= \sum_{x_0} P(x_0) \cdot \rho_{v_Q}(v_Q) \int w(x, v_Q; t, x_0) \cdot dx = \sum_{x_0} P(x_0) \cdot \rho_{v_Q}(v_Q) \\
&= \rho_{v_Q}(v_Q) \sum_{x_0} P(x_0) = \rho_{v_Q}(v_Q).
\end{aligned} \tag{70}$$

# 6 Multidimensional extension

## 6.1 Lattice and particle emissions
The model equations are modified as follows in the $D$-dimensional case, where the lattice is composed of $D$ spatial dimensions $x_d$ and one temporal dimension $t$. Each of the dimensions is characterized by the same fundamental length $X$ and acts independently. However, the value of the fundamental length and time varies with the number of spatial dimensions considered in the picture. In general terms, $m\Delta\tilde{v}^{(N)}\Delta\tilde{x}^{(N)} = 2mcX^{(D)} = 2h/\Omega_D$, where $\Omega_D = 2\pi^{d/2}/\Gamma(d/2)$ is the solid angle in $D$ dimensions. Accordingly, $T^{(D)} = X^{(D)}/c$.

Source preparation fixes the multidimensional source momentum $v_0 = \{v_{0d}\}$, $d = 1, \ldots, D$, and a scalar source phase $\epsilon$.

## 6.2 Microscopic motion

Equation of motion (2) is obviously extended as

$$x_d[n+1] = x_d[n] + v_d[n], \qquad x_d[n_0] = x_{0d}, \tag{71}$$

for $d = 1, \ldots, D$. As for the speed, each $v_d$ is characterized by its own $\boldsymbol{v_d}, \boldsymbol{e_d}$, and (8) is extended as

$$\boldsymbol{v_d}[n] := v_{Qd}[n] + v_{Fd}[n], \qquad \boldsymbol{v_d}[n_0] = v_{0d}. \tag{72}$$

External forces are have $D$ contributions $f_d(x[n'], n')$ that might depend on all of the $D$ positions, whence the argument $x$ without subscripts.

As for the quantum force, equation (12) is extended with some modifications as follows

$$v_{Qd}[n] = \left\{ v_{0d} - \rho_d \sum_i \sum_{j \neq i} v_Q^{(ij)}[n] \right\}, \tag{73}$$

where the function $\{\cdot\}$ has the same meaning as in (10). On the one hand, while in the one-dimensional case the fact that the source momentum is quasi-randomly attributed a value between -1 and 1 enforces that the quantum momentum is also always comprised between these limits, this circumstance is no longer true with more than one dimensions. A general enforcement of the light speed limit requires that each instantaneous violation of such a limit undergoes a correction of the quantum momentum. Consequently also $v_{0d}$ and all boson momenta are corrected.

The quantities $\rho_d \in [0,1]$ are random numbers carried by the particle, such that $\sum_d \rho_d = 1$. They control the split of the boson momenta among the various directions. As such, they are uncorrelated to the source momenta. It is easy to retrieve (12) when $\rho_1 = 1$.

The boson momenta remain scalar quantities and are obtained by extending (19) as

$$v_Q^{(ij)}[n] = \begin{cases} v_Q^{(ij)}[n-1] \cdot \left(1 - \dfrac{1}{2k^{(ij)}[n]}\right), & \text{if } k^{(ij)}[n] > 0 \\ \dfrac{\omega_{\{x\}[n]t[n]}^{(ij)}[n]}{\sum_d^D \rho_d \delta_d^{(ij)}}, & \text{otherwise} \end{cases}, \tag{74}$$

with an obvious meaning of the quantities $\delta_d^{(ij)}$.

Condition $C^{(ij)}$ in (21) and, consequently, all the counters are extended to all dimensions. Conversely, equations (22) and (24) remain scalar, i.e., the quantity $\omega_{xt}^{(ij)}$ is not extended as many times as the number of dimensions. However, (23) is extended as

$$\overline{\omega}_{\xi\tau}^{(ij)}[n] = \begin{cases} \overline{\omega}_{\xi\tau}^{(ij)}[n-1], & \kappa_{\xi\tau}^{(ij)}[n] > 0 \\ \displaystyle\sum_{d=1}^{D} \delta_d^{(ij)} v_{Qd}[n] - \epsilon^{(ij)}, & \text{otherwise} \end{cases}. \tag{75}$$

The $D$-dimensional model is finally summarized as

$$M_D := \begin{cases} t[n] = t[n-1] + 1, & t[n_0] = 0, \\ x_d[n+1] = x_d[n] + v_d[n], & x_d[n_0] = x_{0d}, \\ \Pr(v_d = 0, \pm 1) = (7), \\ \boldsymbol{v_d}[n] := v_{Qd}[n] + v_{Fd}[n], & \boldsymbol{v_d}[n_0] = v_{0d}, \\ v_{0d} = U[-1; 1], \\ v_{Fd}[n] = \sum_{n'=n_0+1}^{n} f_d(x[n'], n'), \\ v_{Qd}[n] = (73), \\ v_{Qd}^{(ij)}[n] = (74) - (75). \end{cases} \quad (76)$$

## 6.3 Probability densities

In the absence of external forces, the procedure in Section 5.2 is still valid to evaluate the position pmf. The steady-state value of the LBM is evaluated as a function of $\overline{\omega}_{\xi\tau}^{(ij)}$,

$$\boldsymbol{\omega}_{\xi\tau}^{(ij)} = \frac{\sin\left(\pi \overline{\omega}_{\xi\tau}^{(ij)}\right)}{\pi}, \quad (77)$$

where

$$\overline{\boldsymbol{\omega}}_{\xi\tau}^{(ij)} = \sum_{d=1}^{D} \delta_d^{(ij)} \frac{\left(\xi_d - \xi_{0d}^{(i)}\right) + \left(\xi_d - \xi_{0d}^{(j)}\right)}{2\tau} - \epsilon^{(ij)} = \sum_{d=1}^{D} \delta_d^{(ij)} \frac{\xi_d - \frac{\xi_{0d}^{(i)} + \xi_{0d}^{(j)}}{2}}{\tau} - \epsilon^{(ij)}. \quad (78)$$

The expected value of $v_Q^{(ij)}$ is now evaluated as

$$\boldsymbol{v}_Q^{(ij)} = \sqrt{P^{(ij)}} \frac{\sin\left(\pi \overline{\omega}_{xt}^{(ij)}\right)}{\pi \sum_d^D \rho_d \delta_d^{(ij)}}. \quad (79)$$

The expected value of the particle's quantum momentum is eventually evaluated as

$$\boldsymbol{v_{Qd}} = v_{0d} - \rho_d \sum_i \sum_{j \neq i} \sqrt{P_0^{(i)} P_0^{(j)}} \frac{\sin\left(\pi \sum_{d=1}^{D} \delta_d^{(ij)} \frac{x_d - \frac{x_{0d}^{(i)} + x_{0d}^{(j)}}{2}}{t} - \pi \epsilon^{(ij)}\right)}{\pi \sum_d^D \rho_d \delta_d^{(ij)}}. \quad (80)$$

Now the joint pmf of the $D$ position coordinates must be evaluated from the definition

$$\rho(\boldsymbol{x_1}, \ldots, \boldsymbol{x_D}; t) = \rho(v_{01}, \ldots, v_{0D}) \cdot \det \begin{bmatrix} \frac{\partial v_{01}}{\partial x_1} & \cdots & \frac{\partial v_{01}}{\partial x_D} \\ \vdots & \ddots & \vdots \\ \frac{\partial v_{0D}}{\partial x_1} & \cdots & \frac{\partial v_{0D}}{\partial x_D} \end{bmatrix}. \quad (81)$$

Observing that $\boldsymbol{x_d} = x_{0d} + \boldsymbol{v_{Qd}} t$ holds, the cross terms cancel out in the determinant of (81), yielding

$$\rho(x_1,...,x_D;t) = \frac{1 + \sum_i \sum_{j \neq i} \sqrt{P_0^{(i)} P_0^{(j)}} \cos\left(\pi \sum_{d=1}^{D} \delta_d^{(ij)} \frac{x_d - \frac{x_{0d}^{(i)} + x_{0d}^{(j)}}{2}}{t} - \pi \epsilon^{(ij)}\right)}{2t}. \tag{82}$$

Again, by recognizing that the cosine argument divided by $\pi$ is equal to

$$\frac{\sum_{d=1}^{D} 2x_d \left(x_{0d}^{(i)} - x_{0d}^{(j)}\right) - \left(\sum_{d=1}^{D} \left(x_{0d}^{(i)}\right)^2 - \sum_{d=1}^{D} \left(x_{0d}^{(j)}\right)^2\right)}{2t} - \epsilon^{(ij)} = -\Delta S^{(ij)}(x,t). \tag{83}$$

the Schrödinger equation and Born rule follows.

The case with external and quantum forces follows analogously. In the following section, some particular scenarios with potential barriers and geometrical constraints are described.

### 6.4 Potential barriers in a 2-d space

In a 2d space a potential barrier can be defined by a region of any form. Consider a barrier defined by a succession of lattice nodes $x \in \mathcal{X}_B$ such that $a_1 x_1 - a_2 x_2 = a_0$. Each time a particle hits the barrier, it gains external bosons

$$f_d(x \in \mathcal{X}_B, n) = -\left(\frac{a_d^2 - a_{d'}^2}{a_d^2 + a_{d'}^2} + 1\right) v_{Qd}[n-1] + \frac{2 a_d a_{d'}}{a_d^2 + a_{d'}^2} v_{Qd'}[n-1], \tag{84}$$

where $d' := \mathrm{mod}(d, 2) + 1$. One retrieves the 1d case with $a_2 = 0$. Consequently, at each barrier hit $v_{Qd}[n] = v_{Qd}[n-1] + f_d[n]$.

The span re-initialisation mechanism that occurs at each barrier hit is modified in accordance:

$$\ell_d[n] = -\frac{a_d^2 - a_{d'}^2}{a_d^2 + a_{d'}^2} \ell_d[n-1] + \frac{2 a_d a_{d'}}{a_d^2 + a_{d'}^2} \ell_{d'}[n-1], \tag{85}$$

Again, one retrieves the 1d case with $a_2 = 0$. Note that both the coefficients $2a_1 a_2/(a_1^2 + a_2^2)$ and $(a_1^2 - a_2^2)/(a_1^2 + a_2^2)$ are rational numbers.

Instead of being constants, variable quantities $a_1$, $a_2$ can be attributed to each single lattice node on the wall, so that each node encodes a pair of external bosons of the form above.

### 6.5 Particle on a line in a 2-d space

In this scenario, the particle is forced to follow a succession of lattice sites ("line") such that $a_1 x_1 - a_2 x_2 = a_0$. On the line $f = 0$, while everywhere else the particle gains an external boson and has its span reinitialized according to the rules (84)-(85). As a consequence of rules (84), the combination of quantum momenta that is conserved after each hit, $v_{Q\parallel}[n] = v_{Q\parallel}[n-1]$, is that parallel to the line, $v_{Q\parallel} := \kappa(a_2 v_{Q1} + a_1 v_{Q2})$, for any measure $\kappa$. On the other hand, the combination perpendicular to the line $v_{Q\perp} := \kappa(a_1 v_{Q1} - a_2 v_{Q2})$ is inverted at each hit due to the external boson gained, $v_{Q\perp}[n] = -v_{Q\perp}[n-1]$.

Note that this scenario is equivalent to a 2d particle-in-a-box scenario, where one dimension has a box width of $2a = 1$. The normal component plays the same role as the quantum momentum in a 1d particle-in-a-box scenario: it ultimately tends to take the stationary values $\frac{n}{a} = 2n$, which is admissible only for $n = 0$ (otherwise it would be superluminal), that is, the null value.

To ensure infraluminal quantum momenta, the measure must be $\kappa = 1/\sqrt{a_1^2 + a_2^2}$. With this measure, the external boson momenta are retrieved as $f_1 = -2a_1\kappa v_\perp$, $f_2 = 2a_2\kappa v_\perp$.

As a consequence of rules (85), the span combination that is accumulated iteration after iteration in spite of the barrier hits is the combination $\ell_\parallel := \kappa(a_2\ell_1 + a_1\ell_2)$, $\ell_\parallel[n] = \ell_\parallel[n-1]$. It is easy to verify that this quantity is proportional to the distance spanned along the line. On the other hand, the perpendicular combination $\ell_\perp := \kappa(a_1\ell_1 - a_2\ell_2)$ is inverted at each hit, $\ell_\perp[n] = -\ell_\perp[n-1]$ and thus it is averagely null. In other terms, no motion accumulation is possible perpendicularly to the line.

Instead of being constants, variable quantities $a_1$, $a_2$ can be attributed to each single lattice node on the line, so that each node encodes a pair of external bosons of the form above. The line is then defined by a succession of nodes, $(x_1, x_2)^{(k)}$, $k = 1, \ldots, N_k$ and the external boson parameters are $a_d^{(k)} = \left(x_{d\prime}^{(k+1)} - x_{d\prime}^{(k-1)}\right)/2$.

Note that, while QM describes this particle-on-a-line scenario with a change from Cartesian to curvilinear coordinates, in the proposed model there is no coordinate change needed nor possible, as the lattice is assumed to be fixed. However, the features of QM are captured as they arise naturally from the lattice description. The particle on a ring scenario shown in the Appendix confirms this feature.

Similar considerations apply for the case of a particle bounded to remain on a surface in a three-dimensional space.

# 7 Extension to two-particle, two-state entanglement

Two-particle systems where each particle evolve independently of the other are naturally represented in the proposed model. It is sufficient to attribute independent source quantities to the two emission sets, in addition to some "flag" that distinguishes the two sets. The Reset Condition is activated only for particles and lattice traces having the same flag.

Conversely, an extension of the model that covers momentum-entangled particles (in one dimension) is proposed in this section.

## 7.1 Particle emissions

Entangled particles are emitted at sources as pairs. The two entangled particles are denoted with superscripts $R = \{I, II\}$. The model being limited to dichotomic systems, we shall assume here a two-source distribution $x_0^{\{a,b\}}$ such as $\left|x_0^{(a)} - x_0^{(b)}\right| = \delta$ and $P_0^{\{a,b\}} = 1/2$. Entangled particles are assumed in this presentation to be emitted from opposite sources (the alternative choice could be made as well, leading to a sign change in the model).

The source preparation attributes "entangled" source momenta, according to the rule

$$v_0^{(II)} + v_0^{(I)} = 0 . \qquad (86)$$

We shall denote, without loss of generality, $v_0^{(I)} = v_0$ and, consequently, $v_0^{(II)} = -v_0$, where $v_0 = U[-1,1]$ as above.

The source phase $\epsilon^{(\{a,b\},R)}$ is allowed to be different for the two particles, in order to represent scenarios where particles experience additional and uncorrelated phases along their respective paths, as it is the case in interferometers.

In addition to source momentum and phase, each particle is attributed a further random quantity $\phi^{(R)} = U[-1,1]$, the same for both particles ($\phi^{(I)} = \phi^{(II)} = \phi$), which affects phase difference. If $\epsilon^{(\delta,R)} := \epsilon^{(a,R)} - \epsilon^{(b,R)}$, the total phase difference is thus $\epsilon^{(\delta,R)} - \phi$.

## 7.2 Microscopic motion

The motion of entangled particles is assumed to be independent. In other terms, particles, say, I do not interact with lattice bosons generated by the particles II, and vice versa.

The general motion rules introduced in the previous sections apply, except for a few changes. First, with $n_R = 2$ entangled particles, the LBM is twice the one that would be created by non-entangled particles, i.e., equation (15) is modified as

$$\bar{\omega}_{\xi\tau}^{(\delta,R)}[n] = n_R \left( \delta v_Q^{(R)}[n] - \varepsilon^{(\delta,R)} + \phi \right). \tag{87}$$

Additionally the PBM is half the one that would be created for non-entangled particles, that is, equation (18) is modified as

$$v_Q^{(\delta,R)}[n] = \frac{1}{n_R} \frac{\omega_{x[n]t[n]}^{(\delta,R)}[n-1]}{\delta}. \tag{88}$$

Finally, (12) is modified to incorporate the phase $\phi$ as

$$v_Q^{(R)}[n] = v_0^{(R)} - \left\{ v_Q^{(\delta,R)}[n] - \frac{\varepsilon^{(\delta,R)} - \phi}{\delta} \right\}, \tag{89}$$

Note that (15), (18), and (12) are retrieved for the non-entangled case by setting $n_R = 1$ and $\phi = 0$ (any source phase then plays no role for the establishment of the position and momentum pdf's).

As simple as they are, these rules prove to be capable to predict QM outcomes of interferometry experiments, as well as capture violations of Bell's inequalities.

## 7.3 Probability densities

The joint pmf of the position of two entangled particles is denoted as $\rho(x^{(I)}, x^{(II)}; t)$ and represents the probability that both particles of the same emission arrive exactly at the indicated locations $x^{(I)}$, resp., $x^{(II)}$, in a time $t$.

Considering a scenario without external forces, $x^{(R)} = v_Q^{(R)} t$, the expected values of the quantum momenta of the entangled particles are obtained similarly to Sect. 5.2 as

$$v_Q^{(R)} = v_0^{(R)} + \frac{\varepsilon^{(\delta,R)} - \phi}{\delta} - \frac{\sin\left(2\left(\pi\delta\frac{x^{(R)}}{t} - \pi\epsilon^{(\delta,R)} + \pi\phi\right)\right)}{2\pi\delta}, \tag{90}$$

where, for the two-state systems considered, the term $\left(x_0^{(a)} + x_0^{(b)}\right)/2t$ vanishes or becomes negligible for large times (thus letting the pdf's be independent of the sources as in (67)). Equations (90) are rewritten as

$$H_I = v_0 - \frac{x^{(I)}}{t} + \frac{\varepsilon^{(I)} - \phi}{\delta} - \frac{\sin\left(2\left(\pi\delta\frac{x^{(I)}}{t} - \pi\epsilon^{(I)} + \pi\phi\right)\right)}{2\pi\delta} = 0, \tag{91}$$

$$H_{II} = -v_0 - \frac{x^{(II)}}{t} + \frac{\varepsilon^{(II)} - \phi}{\delta} - \frac{\sin\left(2\left(\pi\delta\frac{x^{(II)}}{t} - \pi\epsilon^{(II)} + \pi\phi\right)\right)}{2\pi\delta} = 0. \tag{92}$$

The quantities $x^{(I)}$ and $x^{(II)}$ are now functions of two stochastic variables, namely, $v_0$ and $\phi$. The joint position pdf is therefore evaluated from

$$\rho(x^{(I)}, x^{(II)}; t) \propto \rho(v_0, \phi) \cdot \left|\det\begin{bmatrix} \frac{\partial v_0}{\partial x^{(I)}} & \frac{\partial v_0}{\partial x^{(II)}} \\ \frac{\partial \phi}{\partial x^{(I)}} & \frac{\partial \phi}{\partial x^{(II)}} \end{bmatrix}\right|, \tag{93}$$

where the determinant in (93) is evaluated for the values of $\phi$ and $v_0$ that simultaneously fulfil (91)-(92). The partial derivatives in (93) are evaluated as

$$\frac{\partial v_0}{\partial x^{(I)}} = -\frac{\partial H_I}{\partial x^{(I)}} \bigg/ \frac{\partial H_I}{\partial v_0} = \frac{1 + c_I}{t} \tag{94}$$

$$\frac{\partial v_0}{\partial x^{(II)}} = -\frac{\partial H_{II}}{\partial x^{(II)}} \bigg/ \frac{\partial H_{II}}{\partial v_0} = -\frac{1 + c_{II}}{t}$$

$$\frac{\partial \phi}{\partial x^{(I)}} = -\frac{\partial H_I}{\partial x^{(I)}} \bigg/ \frac{\partial H_I}{\partial \phi} = -\frac{\delta}{t}$$

$$\frac{\partial \phi}{\partial x^{(II)}} = -\frac{\partial H_{II}}{\partial x^{(II)}} \bigg/ \frac{\partial H_{II}}{\partial \phi} = -\frac{\delta}{t}$$

where we denote for simplicity $c_I := \cos(2f_I)$ and $c_{II} := \cos(2f_{II})$, with $f_I := \pi\delta x^{(I)}/t - \pi\epsilon^{(\delta,I)} + \pi\phi$ and $f_{II} := \pi\delta x^{(II)}/t - \pi\epsilon^{(\delta,II)} + \pi\phi$. Inserting (94) into (93), we obtain

$$\rho(x^{(I)}, x^{(II)}; t) \propto \frac{1}{2} \cdot \frac{1}{2} \cdot \frac{\delta}{t^2} |-(1 + c_I) - (1 + c_{II})|. \tag{95}$$

Equations (91) and (92) are simultaneously fulfilled when $f_I = -f_{II}$. That leads to the condition

$$\phi = \frac{\epsilon^{(\delta,II)} + \epsilon^{(\delta,I)}}{2} - \delta\frac{x^{(I)} + x^{(II)}}{2t}, \tag{96}$$

and thus to

$$c_I = \cos\left(\pi(\epsilon^{(\delta,II)} - \epsilon^{(\delta,I)}) + \pi\delta\frac{x^{(I)} - x^{(II)}}{t}\right),$$

$$c_{II} = \cos\left(-\pi(\epsilon^{(\delta,II)} - \epsilon^{(\delta,I)}) - \pi\delta\frac{x^{(I)} - x^{(II)}}{t}\right) = c_I. \tag{97}$$

Finally, the joint pdf is evaluated inserting (97) in (95) and normalizing, as

$$\rho(x^{(I)}, x^{(II)}; t) = \frac{1}{4t^2}\left(1 + \cos\left(\pi(\epsilon^{(\delta,II)} - \epsilon^{(\delta,I)}) + \pi\delta\frac{x^{(I)} - x^{(II)}}{t}\right)\right). \tag{98}$$

A further analysis of (45) remarkably allows to retrieve the Born rule for two-particle entangled systems. Recognize that the cosine argument divided by $\pi$ is equal to

$$\left(\epsilon^{(\delta,II)} - \epsilon^{(\delta,I)}\right) + \delta\frac{x^{(I)} - x^{(II)}}{t} = S^{(a,I)} + S^{(b,II)} - S_I^{(b,I)} - S_{II}^{(a,II)}. \tag{99}$$

The right-hand side of (45) can be then equivalently obtained as the phase of a complex number $\psi$ defined as

$$\psi = \psi^{(a,I)}\psi^{(b,II)} + \psi^{(b,I)}\psi^{(a,II)}, \tag{100}$$

where each term of (100) is the probability amplitude of the free particle emitted by source $a$ or $b$. Equation (100) is clearly how QM evaluates the two-particle amplitude in the case of entanglement [], where the amplitude is not factorizable as a product of two independent terms. With this observation, the equivalence between the proposed model and quantum mechanics is demonstrated for two-particle, two-state entanglement scenarios.

Note that, although $\phi$ plays the role of a "hidden variable" as those explicitly discarded by Bell's theorem, the pdf (93) is not in the form postulated by this theorem. In fact, (93) is generally not factorized as the product of two (bivalued) terms, separately depending on $\epsilon^{(\delta,I)}$ and $\epsilon^{(\delta,II)}$ (a Bell's assumption often referred to as "setting independence" or "no-signalling" [39]) albeit possibly from a common set of hidden variables. Indeed, in (97), $c_I$ and $c_{II}$ depend on both $\epsilon^{(\delta,I)}$ and $\epsilon^{(\delta,II)}$ via the source phase $\phi$ given by (96). Therefore, the proposed model is not forbidden by Bell's theorem, which is based on Bell's assumptions, to violate Bell's inequalities.

Nevertheless, nowhere in the proposed model, particles, say, II "know" about which phase difference experience particles I, thus locality still applies even if the "setting independence" assumption is not valid. The emergence of a particular phase $\phi$ that is a function of both $\epsilon^{(\delta,I)}$ and $\epsilon^{(\delta,II)}$ results from the fact that for other values of $\phi$, the desired combination of $x^{(I)}$ and $x^{(II)}$ is impossible or, at least, highly improbable. In this respect, the key feature of the proposed model is the dependency of the particle trajectories on two random "hidden variables".

# 8 Numerical simulation

This section describes the numerical codes used to simulate QM scenarios with the proposed model. As a general feature of the simulations presented in next section, the model equations are repeated for a series of $N_p$ consecutive particle emissions, so that the motion of each particle is simulated for $N_t$ iterations. Probability mass function of position is retrieved as the frequency of arrivals.

The stabilisation of the quantum mechanisms proposed and the emergence of quantum-like behaviour require a large number of emissions $N_p$ (and large times $N_t$). In order to speed up the calculations and make their reasonable for personal computers, it is assumed that (i) the lattice is already trained after a large number of previous, non-simulated emissions, and (ii) the particle is also trained.

As a general feature of the simulations presented in next section, the model equations are repeated for a series of $N_p$ consecutive particle emissions, so that the motion of each particle is simulated for $N_t$ iterations. Probability mass function of position is retrieved as the frequency of arrivals.

The stabilisation of the quantum mechanisms proposed and the emergence of quantum-like behaviour require a large number of emissions $N_p$ (and large times $N_t$). In order to speed up the calculations and make their reasonable for personal computers, it is assumed that (i) the lattice is already trained after a large number of previous, non-simulated emissions, and (ii) the particle is also trained.

## 8.1 Accelerated codes

### 8.1.1 Trained code

To dispose of an accelerated code that converges faster to quantum mechanical results, while taking into account the fact that lattice and particle "training" are not instantaneous processes, an artificial time lag of a few $n_\tau$ iterations is introduced. Additionally, the terms $\frac{x_0^{(i)}+x_0^{(j)}}{2}$ (reflecting the source locations stored at the lattice nodes visited by the particle) are replaced by the true particle source location $x_0$ as in (67). The model (19)-(24) is thus replaced by its "trained" counterpart

$$v_Q^{(ij)}[n] = \frac{\sqrt{P^{(ij)}} \sin\left(\pi(\delta^{(ij)}q[n] - \epsilon^{(ij)})\right)}{\pi \delta^{(ij)}}, \quad (101)$$
$$q[n] = q[n-1] \cdot \frac{(n_\tau - 1)}{n_\tau} + \frac{1}{n_\tau} \cdot v_Q[n].$$

with the $P^{(ij)}$, $\delta^{(ij)}$, and $\epsilon^{(ij)}$ pre-computed for each $ij$ pairs. In summary, the accelerated code has the following subjacent motion model:

$$M^* := \begin{cases} t[n] = t[n-1] + 1, & t[n_0] = 0, \\ x[n+1] = x[n] + v[n], & x[n_0] = x_0, \\ \Pr(v = 0, \pm 1) = (7), \\ v[n] := v_Q[n] + v_F[n], & v[n_0] = v_0, \\ v_0 = U[-1;1], \\ v_F[n] = \sum_{n'=n_0+1}^{n} f(x[n'], n'), \\ v_Q[n] = v_0 - \sum_i \sum_{j \neq i} v_Q^{(ij)}[n], \\ v_Q^{(ij)}[n] = \frac{\sqrt{P^{(ij)}} \sin\left(\pi(\delta^{(ij)}q[n] - \epsilon^{(ij)})\right)}{\pi \delta^{(ij)}} \\ q[n] = q[n-1] \cdot \frac{(n_\tau - 1)}{n_\tau} + \frac{1}{n_\tau} \cdot v_Q[n] \end{cases} \quad (102)$$

### 8.1.2 Expected-values code

Additional reduction of computing times are achieved with an ultra-accelerated code that directly simulates the expected values of both position ($x$) and the quantum momentum ($v_Q$). The subjacent motion model $M^{**}$ can be summarized as follows, with time dependency replacing iterations:

$$M^{**} := \begin{cases} x(t) = A(t)x_0 + B(t)v_Q + C(t), & x(0) = x_0 \\ v_Q - v_0 + \sum_i \sum_{j \neq i} \frac{\sqrt{P^{(ij)}} \sin\left(\pi(\delta^{(ij)}v_Q - \epsilon^{(ij)})\right)}{\pi \delta^{(ij)}} = 0, \\ v_0 = U[-1;1] \end{cases} \quad (103)$$

This ultra-accelerated code directly integrates the classical equation of motion in the presence of external forces, albeit introducing a randomness in the source momentum. As such, it does not require to simulate explicitly either the particle momentum or its expected value.

The question might even arise if the discrete nature of model $M$ and the particle-lattice interaction proposed is really necessary in the proposed ontology. In other terms, if the quasi-deterministic, continuous-space model $M^{**}$ would be fundamental enough.

However, model $M^{**}$ does not describe the penetration of particles in classically forbidden regions of space. For instance, scenario 9.5 (Delta potential) and in particular the Gaussian-wave source preparation could not be properly simulated with this model: particles would tunnel through the potential barrier only for $v_Q^2$ large enough, contrarily to what expected.

Moreover, model $M^{**}$ is nonlocal, in the sense that a particle emitted at a certain source would know about other possible sources since the beginning of its evolution (leading to the circumstance that $\boldsymbol{v}_Q$ is determined at the preparation, see (103), which is in contrast with the requirements set in the Introduction.

These facts reveals that integer quantities and in particular discrete spacetime are necessary in $M$ for at least two reasons: (i) to establish a local exchange between particles and the lattice and establish quantum forces, and (ii) to introduce a certain probability for particles to tunnel through potential barriers regardless of their momentum. The arithmetic operations on the these integer quantities emerge mostly from probability rules or counter updates.

The emergence of the square root in the formula for the average boson momentum, as well as that of the sine function in the formula for the steady-state footprint momentum can only derive from a discrete formulation of the respective decay laws. Additionally, the encounter between a flying particle and a footprint can only occur precisely if discrete space and time are assumed. These two observations motivate the existence of a discrete spacetime lattice in the model detailed in the following sections.

<div align="center">***</div>

The next sections describe the numerical experiments that have been performed to show that predictions of the model are consistent with QM. Results of simulations of several one- and multi-dimensional scenarios are presented in the Appendix A. One-dimensional scenarios comprise of free particle (319.1), constant force (9.2), harmonic oscillator (9.3), particle in a box (9.4) and Delta potential (9.5). Multi-dimensional scenarios comprise of free particle (10.1), particle on a ring (10.2) and particle on a sphere (10.3).

For each of these cases, various QM initial states $\psi(x_0)$ are reproduced (single source, multiple equally separated sources, Gaussian waves, stationary states, etc.). Generally, the source probability is taken as $P_0^{(k)} = P_0(x_0 = k) = |\psi(x_0)|^2$, while the source phase (required, for instance, for propagating Gaussian waves) is taken as $\epsilon^{(k)} = \epsilon(x_0 = k) = \angle\psi(x_0)/\pi$, the angle of the QM initial amplitude divided by $\pi$.

Results of the proposed model are compared with quantum mechanical predictions, computed by applying the respective amplitude propagators to the initial states selected.

| § | Scenario | Parameters | no. of sources $N_s$ | no. of bosons $N_b/2$ |
|---|---|---|---|---|
| 9.1 | Free particle | | | |
| | Single source | | 1 | 0 |

|     | Two sources                  | $D = 1$                                              | 2        | 1      |
|-----|------------------------------|------------------------------------------------------|----------|--------|
|     | Multiple sources             | $a = 3$                                              | 3        | 3      |
|     | Plane wave (stationary state)| $\ell = 100, v_\varphi = 0.1$                        | 101      | 5050   |
|     | Gaussian wave                | $a = 0, v_\varphi = 0.1, d = 5$                      | 31       | 465    |
| 9.2 | Free faller                  |                                                      |          |        |
|     | Single source                | $\phi = 2 \cdot 10^{-3}$                             | 1        | 0      |
|     | Gaussian wave                | $\phi = 2 \cdot 10^{-3}, a = 0, v_\varphi = 0.1, d = 5$ | 31    | 465    |
| 9.3 | Harmonic oscillator          |                                                      |          |        |
|     | Single source                | $\Omega = .005$                                      | 1        | 0      |
|     | Gaussian wave                | $\Omega = 1 \cdot 10^{-4}, a = 0, v_\varphi = 0.1, d = 5$ | 27   | 351    |
|     | Stationary state             | $\Omega = 1 \cdot 10^{-4}, n = 1$                    | 157      | 12246  |
| 9.4 | Particle in a box            |                                                      |          |        |
|     | Single source                | $a = 10$                                             | 1 (21)   | 210    |
|     | Stationary state             | $a = 10, n = 3, 5$                                   | 21 (315) | 49455  |
| 9.5 | Delta potential              |                                                      |          |        |
|     | Single source                | $x_0 = 10, \ell = 15$                                | 1        | 0*     |
|     | Gaussian wave                | $a = 10, d = 10, v_\varphi = -0.3, \ell = 2$         | 1        | 0*     |
| 10.1| 2-dim. free particle         |                                                      |          |        |
|     | Single source                |                                                      | 1        | 0      |
|     | Multiple sources             | $a = 2, b = 1$                                       | 3        | 3      |
| 10.2| Particle on a ring           |                                                      |          |        |
|     | Single source                | Closed-form                                          |          |        |
|     | Plane wave                   | $r = 10, m = 4$                                      | 64 (1311)| 858705 |
|     | Stationary state             | $r = 10, n = 1$                                      | 64 (1311)| 858705 |
| 10.3| Particle on a sphere         |                                                      |          |        |
|     | Single source                | Closed-form                                          |          |        |
|     | Stationary state             | $r = 100, \ell = 4, m = 0$                           | 314(×628)| 49141  |

# 9 Numerical results: One-dimensional scenarios

Simulations have been carried on with the code $M^*$, with both the lattice and the particles trained ($n_\tau = t$). The scenarios considered here, namely, (1) free particle, (2) "free faller" (particle submitted to a constant force), (3) harmonic oscillator, (4) "particle in a box", and (5) Delta potential allow observing the emergence of interference, quantization of momentum and energy, as well as tunnelling in the proposed model.

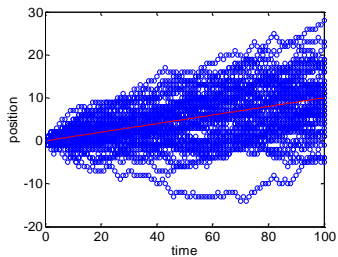
(a)

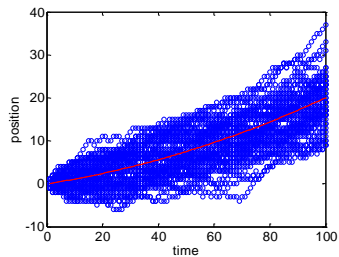
(b)

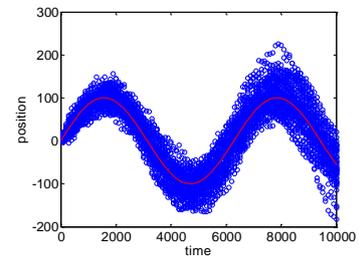
(c)

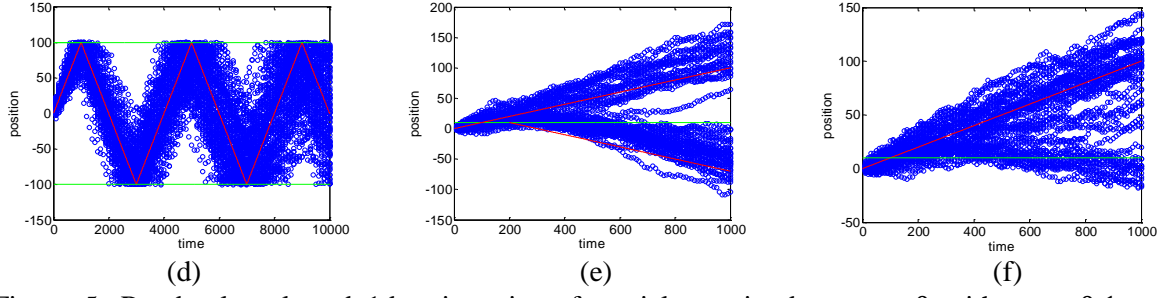

Figure 5: Randomly selected 1d trajectories of particles emitted at $x_0 = 0$ with $v_0 = 0.1$ under several force scenarios: (a) free particle; (b) free faller; (c) harmonic oscillator; (d) particle in a box; (e) Delta potential larger than kinetic energy; (f) Delta potential smaller than kinetic energy. Red curves are the classical/expected trajectories. Green curves are potential barriers.

## 9.1 Free particle

For all the scenarios in this section, $f \equiv 0$. The expected motion is $\mathbf{x} = x_0 + \mathbf{v_0}t$. Example trajectories of particles emitted from the same source with the same source momentum are plotted in Figure 5a.

Ensemble results are compared with those of quantum mechanics (theoretical values) obtained by using the propagator

$$K^{(FP)}(x,t|x_0) = \frac{1}{\sqrt{2\iota t}} \cdot \exp\left(\frac{\iota \pi (x-x_0)^2}{2t}\right) \tag{104}$$

in lattice units. Results for single-source, two-source (double slit), multiple-source, Gaussian wave, and plane-wave (stationary state) preparations are shown.

### 9.1.1 Single source

In this case, $x_0 = \{x_0\}$, $P_0 = \{1\}$. In Sect. 3.1 the equivalence between the proposed model and quantum mechanics has been already demonstrated while obtaining (31), i.e.

$$\rho(\mathbf{x};t) = \frac{1}{2t} \Longrightarrow \mathbf{x} = U[x_0 - t, x_0 + t]. \tag{105}$$

Figure 6 and Figure 7 show the frequency of arrivals after $N_t = 500$ iterations, obtained with an increasing number of emissions at $x_0 = 0$. As $N_p$ increases, the frequency clearly tends to the theoretical probability density (105).

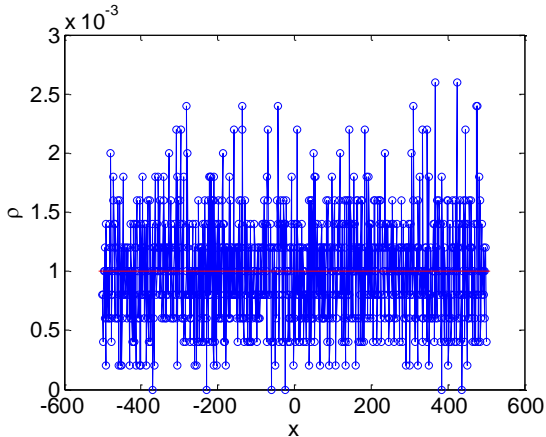

Figure 6: Arrival frequency (blue) and theoretical value (red) for $N_p = 5000$, $t = 500$ as a function of

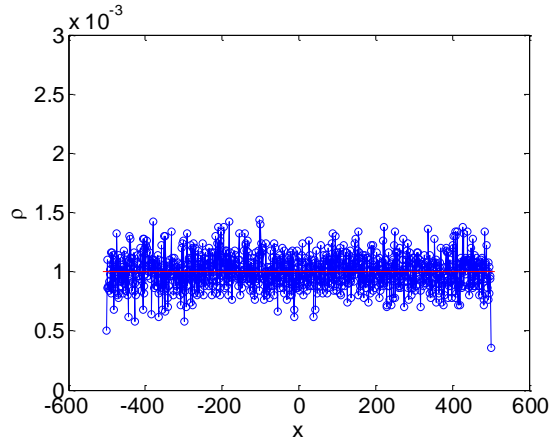

Figure 7: Arrival frequency (blue) and theoretical value (red) for $N_p = 50000$, $t = 500$ as a function of position (free particle, single source, $x_0 = 0$).

position (free particle, single source, $x_0 = 0$).

### 9.1.2 Two sources

This case is equivalent to the classical two-slit experiment, whereas $x_0 = \{-D, D\}$, $P_0 = \{P_0^{(-D)}, P_0^{(D)}\}$, $\epsilon^{(D)} = \epsilon^{(-D)} = 0$, with $P_0^{(D)} + P_0^{(-D)} = 1$. Consequently, two types of boson arise, with $\delta^{(12)} = \delta^{(21)} = 2D$, and $\epsilon^{(12)} = \epsilon^{(21)} = 0$. The theoretically expected pdf is

$$\rho(x; t) = \frac{1 + 2\sqrt{P_0^{(D)} P_0^{(-D)}} \cos\left(2\pi D \frac{x}{t}\right)}{2t}. \tag{106}$$

Figure 10 and Figure 11 show the frequency of arrivals after $N_t = 500$ iterations with emissions at $\pm D = \pm 1$ and with two different sets of source probabilities, $P_0^{(D)} = 1/2$ and $P_0^{(D)} = 0.1$, respectively. In both cases, the frequency clearly tends to the theoretical probability density (106).

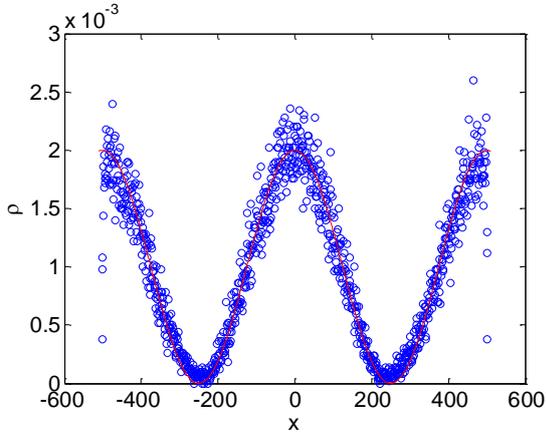
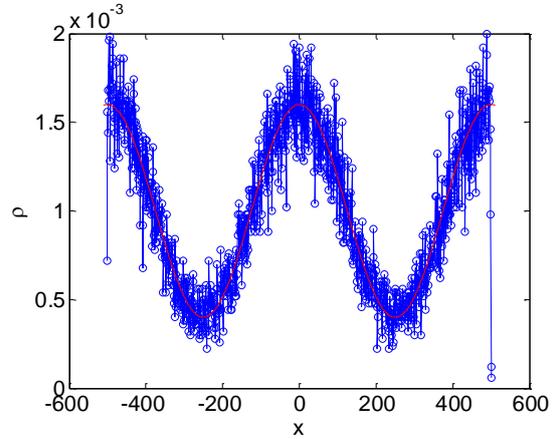

Figure 8: Arrival frequency (blue) and theoretical value (red) for $N_p = 50000$, $t = 500$ as a function of position (free particle, two sources, $D = 1$, $P_0^{(D)} = 0.5$).

Figure 9: Arrival frequency (blue) and theoretical value (red) for $N_p = 50000$, $t = 500$ as a function of position (free particle, two sources, $D = 1$, $P_0^{(D)} = 0.1$).

### 9.1.3 Multiple sources diffraction

A generalisation of the previous case to a scenario with an even number $N_s$ of sources separated by $a$ nodes is formalised as $x_0 = \{ka\}$, with $k = -\frac{N_s-1}{2}, \ldots, \frac{N_s-1}{2}$. The source probabilities are such that $\sum_{k=-\frac{N_s-1}{2}}^{\frac{N_s-1}{2}} P_0^{(k)} = 1$ and the source phase is $\epsilon^{(k)} \equiv 0$. In this case $N_s(N_s - 1)$ types of bosons arise, with $\delta^{(ij)} = |i - j|a$ and $\epsilon^{(ij)} = 0$. The theoretically expected pdf is thus

$$\rho(x; t) = \frac{1 + \sum_i \sum_{i \neq i} \sqrt{P_0^{(i)} P_0^{(j)}} \cos\left(\pi \delta^{(ij)} \frac{x - \frac{a(i+j)}{2}}{t}\right)}{2t}. \tag{107}$$

Figure 10 show the frequency of arrivals after $N_t = 500$ iterations with $a = 2$, $N_s = 3$, and $P_0^{(j)} = 1/N_s$, $\forall j$. The frequency clearly tends to the theoretical probability density (107). Figure 11 shows an additional result, namely, the quantum momentum $v_Q$ as a function of the source momentum $v_0$. The momentum distribution peaks at values $\hat{v}^{(n)} = \frac{2n}{a}$, with $n = 0, \pm 1, \ldots$ It is clearly this pattern that builds (107) via the chain rule (28).

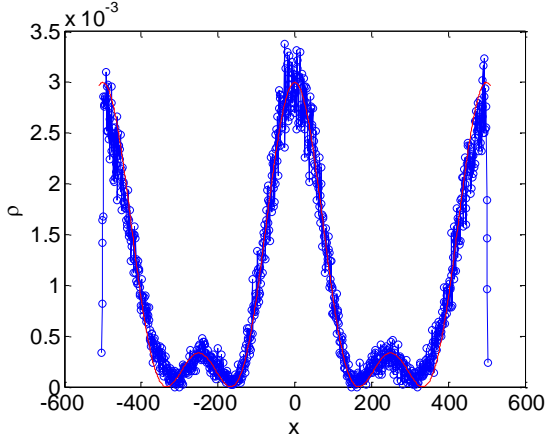
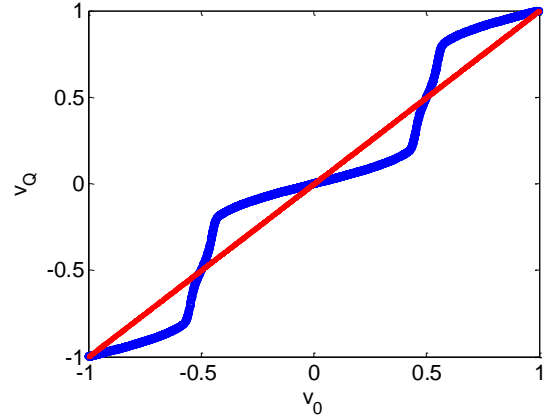

Figure 10: Arrival frequency (blue) and theoretical value (red) for $N_p = 50000$, $t = 500$ as a function of position (free particle, multiple sources, $a = 2$, $N_s = 3$).

Figure 11: Quantum momentum as a function of source momentum for $N_p = 50000$, $t = 500$ (free particle, multiple sources, $a = 2$, $N_s = 3$).

### 9.1.4 Plane wave (stationary state)

A scenario that resembles a "plane wave" state is obtained by placing $\ell + 1$ equiprobable sources separated by one lattice node ($x_0 = \{k\}$, $k \in [-\ell/2, \ell/2]$, $P_0^{(k)} = 1/(\ell + 1)$), and providing these sources with a phase $\epsilon^{(k)} = k v_\varphi$. As this plane wave is a stationary state for the free quantum particle, the theoretically expected pdf is

$$\rho(x; t) \approx \frac{1}{\ell + 1}, \quad x \in [-\ell/2 + v_\varphi t, \ell/2 + v_\varphi t]. \tag{108}$$

for sufficiently large $\ell$; however, for finite $\ell$ the theoretical value results from the application of the propagator (101) to a state $\psi_0 = e^{\iota \pi v_\varphi x}/\sqrt{\ell + 1}$ and is generally different from (108).

Figure 12 shows the frequency of arrival after $N_t = 500$ iterations with $\ell = 100$ and $v_\varphi = 0.1$.

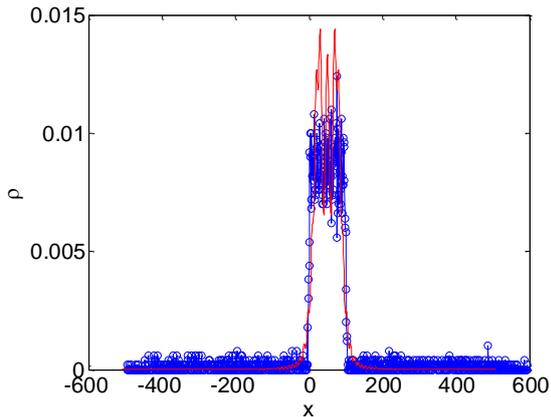
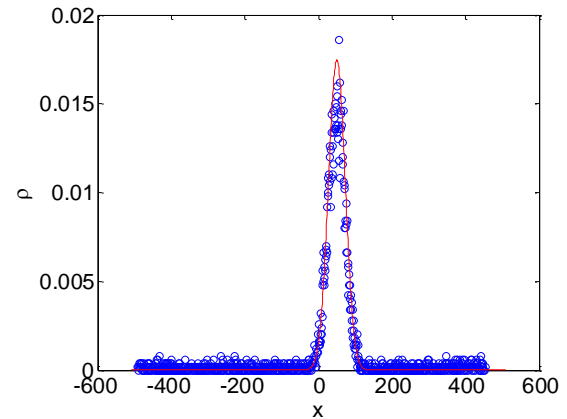

Figure 12: Arrival frequency (blue) and theoretical value (red) for $N_p = 5000$, $t = 500$ as a function of position (free particle, plane wave, $\ell = 100$, $v_\varphi = 0.1$).

Figure 13: Arrival frequency (blue) and theoretical value (red) for $N_p = 5000$, $t = 500$ as a function of position (free particle, Gaussian wave, $a = 0$, $d = 5$, $v_\varphi = 0.1$).

### 9.1.5 Gaussian wave

This scenario implies several sources with different probabilities and phase, which are set to represent the QM initial state $\psi_0(x_0) = (1/\pi d^2)^{1/4} \exp\left(-(x_0 - a)^2/d^2 + \iota\pi v_\varphi(x_0 - a)\right)$. Namely, $x_0 = \{k\}$, $k \in (-\ell, \ell)$, with $P_0^{(k)} = \left(\frac{1}{\pi d^2}\right)^{\frac{1}{2}} \exp\left(-\frac{(k-a)^2}{d^2}\right)$ and $\epsilon^{(k)} = (k-a)v_\varphi$. The theoretically expected pdf is obtained by applying the propagator (101) to $\psi_0$ an and is evaluated as

$$\rho(x; t) = \left(\frac{1}{\pi d^2 \left(1 + \frac{t^2}{\pi^2 d^4}\right)}\right)^{\frac{1}{2}} \exp\left(-\frac{(x - a - v_\varphi t)^2}{d^2 \left(1 + \frac{t^2}{\pi^2 d^4}\right)}\right). \tag{109}$$

Figure 13 shows the frequency of arrival after $N_t = 500$ iterations with $a = 0$, $v_\varphi = 0.1$, and $d = 5$. With these data, in practise, only 31 sources have non-negligible source probability.

## 9.2 Free faller

For all the scenarios in this section, $f(x) \equiv \phi$. The expected motion is $x = x_0 + v_Q t + \phi t^2/2$. Example trajectories of particles emitted from the same source with the same source momentum are plotted in Figure 5b.

Ensemble results are compared with those of quantum mechanics (theoretical values) obtained by using the propagator

$$K^{(FF)}(x, t|x_0) = \frac{1}{\sqrt{2\iota t}} \exp\left(\frac{\iota\pi(x - x_0)^2}{2t} + \frac{\iota\pi(x + x_0)\phi t}{2} - \frac{\iota\pi\phi^2 t^3}{24}\right), \tag{110}$$

where the exponential argument is clearly the classical action (in lattice units) multiplied by the factor $\iota\pi$. Results for single-source and Gaussian-wave preparations are shown.

### 9.2.1 Single source

In this case, $x_0 = \{x_0\}$, $P_0 = \{1\}$. The theoretically expected pdf is

$$\rho(x; t) = \frac{1}{2t}, \quad x \in \left[x_0 - t + \frac{\phi t^2}{2}, x_0 + t + \frac{\phi t^2}{2}\right]. \tag{111}$$

Figure 14 shows the frequency of arrival after $N_t = 200$ iterations with $\phi = 2 \cdot 10^{-3}$ and $N_p = 50000$. Arrivals clearly tend toward distribution (111), with $x \in [-160, 240]$.

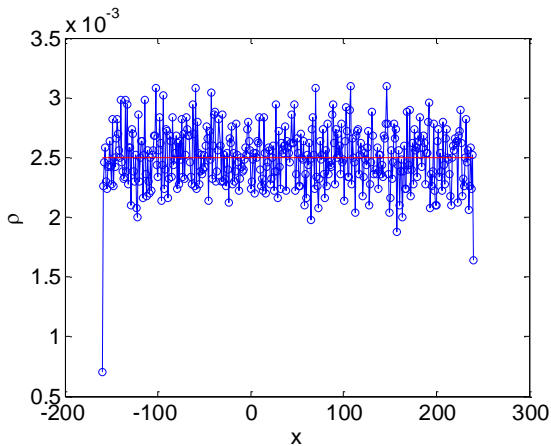
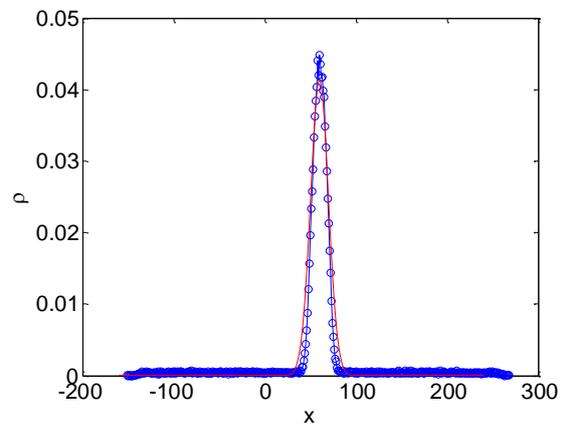

Figure 14: Arrival frequency (blue) and theoretical   Figure 15: Arrival frequency (blue) and theoretical

value (red) for $N_p = 50000$, $t = 200$ as a function of position (free faller, single source, $x_0 = 0$, $\phi = .002$).

value (red) for $N_p = 50000$, $t = 200$ as a function of position (free faller, Gaussian wave, $\phi = .002$, $a = 0$, $d = 5$, $v_\varphi = 0.1$).

### 9.2.2 Gaussian wave

In this case the sources are prepared as in Sect. 9.1.e. The theoretically expected pdf is found by applying the propagator (110) to the initial state $\psi_0$ shown in that section and is

$$\rho(x;t) = \left(\frac{1}{\pi d^2 \left(1 + \frac{t^2}{\pi^2 d^4}\right)}\right)^{\frac{1}{2}} \exp\left(-\frac{\left(x - a - v_\varphi t - \frac{\phi t^2}{2}\right)^2}{d^2 \left(1 + \frac{t^2}{\pi^2 d^4}\right)}\right). \tag{112}$$

Figure 15 shows the frequency of arrival after $N_t = 200$ iterations with $\phi = 2 \cdot 10^{-3}$, $a = 0$, $v_\varphi = 0.1$, $d = 5$, and $N_p = 50000$. The centre of the Gaussian wave, initially at $x = 0$, has moved to the left to the point $v_\varphi t + \phi t^2/2 = 60$.

## 9.3 Harmonic oscillator

This scenario is described by distributing external bosons at each node of the lattice, all of them having a momentum $f(x) = -\Omega^2 x$. The expected motion is $x = x_0 \cos \Omega t + v_Q \sin \Omega t / \Omega$. Example trajectories of particles emitted from the same source with the same source momentum are plotted in Figure 5c.

Ensemble results are compared with those of quantum mechanics (theoretical values) obtained by using the propagator

$$K^{(HO)}(x, t|x_0) = \sqrt{\frac{\Omega}{2\iota \sin \Omega t}} \exp\left(\frac{\iota \pi \Omega}{2 \sin \Omega t}\left((x^2 + x_0^2) \cos \Omega t - 2xy\right)\right) \tag{113}$$

where again (quadratic Lagrangian) the argument of the exponential is clearly the classical action multiplied by the factor $\iota \pi$. Results for single-source, stationary-state, and Gaussian-wave preparations are shown.

### 9.3.1 Single source

In this case, $x_0 = \{x_0\}$, $P_0 = \{1\}$. The theoretically expected pdf is

$$\rho(x;t) = \frac{\Omega}{2|\sin(\Omega t)|},$$
$$x \in \left[x_0 \cos(\Omega t) - \frac{1}{\Omega}\sin(\Omega t), x_0 \cos(\Omega t) + \frac{1}{\Omega}\sin(\Omega t)\right]. \tag{114}$$

Figure 16 shows the frequency of arrival after $N_t = 200$ iterations with $\Omega = .005$ and $N_p = 50000$. Arrivals clearly tend toward distribution (114), with $x \in [-168,168]$.

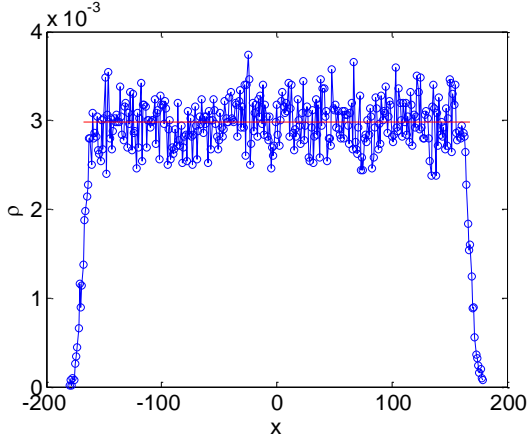
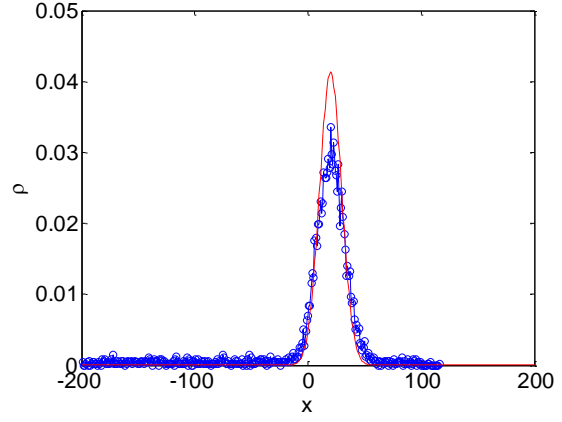

Figure 16: Arrival frequency (blue) and theoretical value (red) for $N_p = 50000$, $t = 200$ as a function of position (harmonic oscillator, single source, $\Omega = .005$, $x_0 = 0$).

Figure 17: Arrival frequency (blue) and theoretical value (red) for $N_p = 5000$, $t = 200$ as a function of position (harmonic oscillator, Gaussian wave, $\Omega = 1 \cdot 10^{-4}$, $a = 0$, $d = 5$, $v_\varphi = 0.1$).

#### 9.3.2 Gaussian wave

In this case the sources are prepared as in Sect. 9.1.e. The theoretically expected pdf is found by applying the propagator (113) to the initial state $\psi_0$ shown in that section and is

$$\rho(\boldsymbol{x};t) \approx \frac{1}{\sqrt{\pi}\sqrt{d^2 \cos^2 \Omega t + \left(\frac{1}{\pi\Omega d}\right)^2 \sin^2 \Omega t}} \exp\left(-\frac{\left(x - a - \frac{v_\varphi \sin \Omega t}{\Omega}\right)^2}{d^2 \cos^2 \Omega t + \left(\frac{1}{\pi\Omega d}\right)^2 \sin^2 \Omega t}\right), \quad (115)$$

Figure 17 shows the frequency of arrival after $N_t = 200$ iterations with $\Omega = 1 \cdot 10^{-4}$, $a = 0$, $v_\varphi = 0.1$, $d = 5$, and $N_p = 5000$. The centre of the Gaussian wave, initially at $x = 0$, has moved to the left to the point $v_\varphi \sin \Omega t / \Omega = 20$.

#### 9.3.3 Stationary states

In this case the source probability and phase are set to represent QM initial states $\psi_0^{(n)}(x_0) = \Omega^{1/4} \frac{1}{\sqrt{2^n n!}} H_n(\sqrt{\pi\Omega} x_0) \exp\left(-\frac{\pi\Omega x_0^2}{2}\right)$, where $H_n$ are the Hermite polynomials. Since these are stationary states, the theoretically expected pdf is obtained as

$$\rho(\boldsymbol{x};t) = P_0^{(x)} = \Omega^{1/2} \frac{1}{2^n n!} H_n(\sqrt{\pi\Omega} x)^2 \exp(-\pi\Omega x^2), \quad (116)$$

that is, a function with $n + 1$ symmetric peaks at some positions $\hat{x}^{(m)}$. Figure 18 shows the frequency of arrival after $N_t = 200$ iterations with $\Omega = 1 \cdot 10^{-4}$, $n = 1$, and $N_p = 5000$.

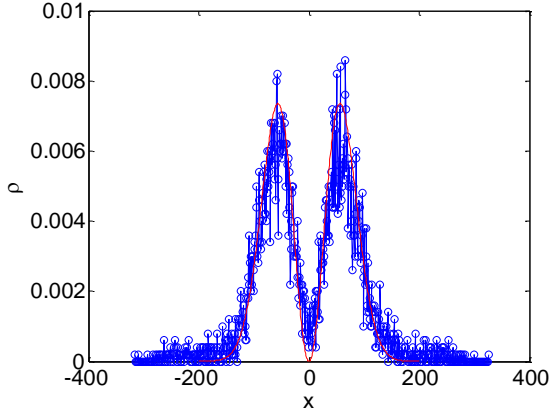
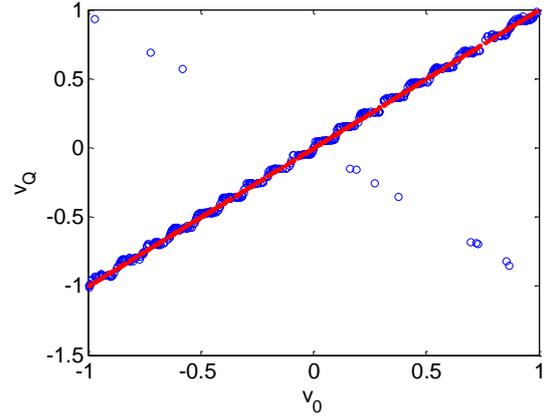

Figure 18: Arrival frequency (blue) and theoretical value (red) for $N_p = 5000$, $t = 200$ as a function of position (harmonic oscillator, stationary state, $\Omega = .0001$, $n = 1$).

Figure 19: Quantum momentum distribution for $N_p = 500$, $t = 500$ as a function of source momentum (particle in a box, single source, $a = 10$, $x_0 = 0$).

The theoretical momentum pdf is

$$\rho_{v_Q}(v_Q) = \frac{1}{2^n n!} \frac{1}{\sqrt{\Omega}} \exp\left(-\frac{\pi v_Q^2}{\Omega}\right) H_n^2\left(\sqrt{\frac{\pi}{\Omega}} v_Q\right), \tag{117}$$

a function that peaks at $\hat{v}_Q^{(m)} = \Omega \hat{x}^{(m)}$. Figure 20 to Figure 22 compare the momentum pdf with the numerical frequency of $v_Q$ for $n = 1$, 2, and 4. Clearly, the two distributions tend to match.

Figure 23 shows the distribution of the quantity $\frac{1}{2}(\Omega^2 x^2 + v_Q^2) = v_Q^2$ (expected particle energy) for $n = 5$ together with the eigenvalue predicted by QM, $E_n = \left(n + \frac{1}{2}\right)\frac{\Omega}{\pi}$ (in lattice units). The figure illustrates that the eigenvalue appears to coincide with the ensemble average value of the expected particle energy.

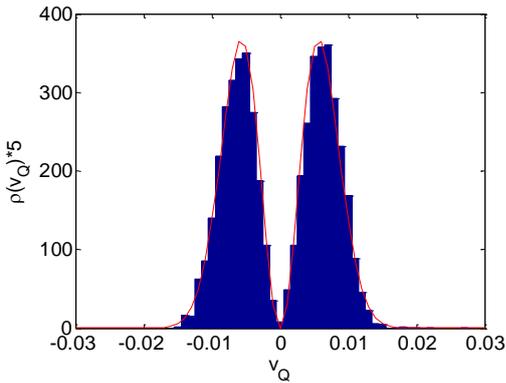
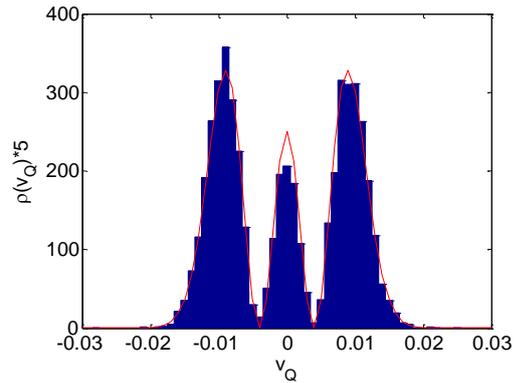

Figure 20: Momentum probability distribution (blue) and non-normalized theoretical value (red) for $N_p = 5000$, $t = 200$ (harmonic oscillator, stationary state, $\Omega = 1 \cdot 10^{-4}$, $n = 1$).

Figure 21: Momentum probability distribution (blue) and non-normalized theoretical value (red) for $N_p = 5000$, $t = 200$ (harmonic oscillator, stationary state, $\Omega = 1 \cdot 10^{-4}$, $n = 2$).

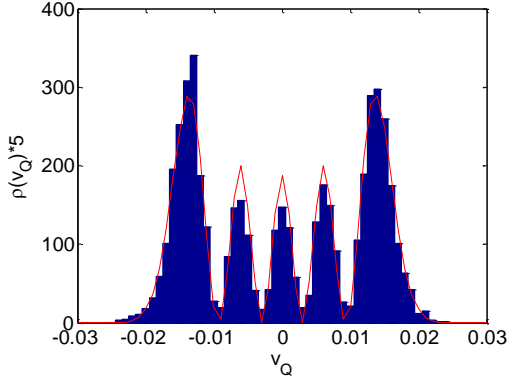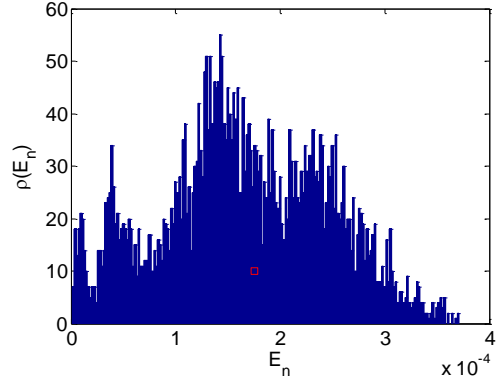

Figure 22: Momentum probability distribution (blue) and non-normalized theoretical value (red) for $N_p = 5000$, $t = 200$ (harmonic oscillator, stationary state, $\Omega = 1 \cdot 10^{-4}$, $n = 4$).

Figure 23: Energy probability distribution (blue) and theoretical expected value (red) for $N_p = 5000$, $t = 200$ (harmonic oscillator, stationary state, $\Omega = 1 \cdot 10^{-4}$, $n = 5$).

### 9.4 Particle in a box

This scenario is defined by the box size $a$, such that $x \in [-a, a]$, with $f = 0$. Each time a particle hits the box boundaries, it gains an external boson $v_F = -2v_Q$ (infinite potential outside the box). Moreover, quantum forces arise even in the presence of a single source because of span sign inversion (13) that occurs at each boundary hit. Equation (50) is replaced by

$$\ell[n] = x[n] - (-1)^h x_0 \pm 2ha, \tag{118}$$

where $h$ is the number of hits and the sign of the last term in the numerator depends on the direction of the first hit. A given node can be thus visited by particles having the same lifetime but different numbers of hits, thus a different span.

The expected motion is $x = \frac{v_Q}{|v_Q|}\mathrm{tri}\left(a, \frac{4a}{|v_Q|}; t + \frac{x_0}{v_Q}\right)$, where $\mathrm{tri}(a', p'; t')$ is the triangular wave with amplitude $a'$ and period $p'$ evaluated at time $t'$. Example trajectories of particles emitted from the same source with the same source momentum are plotted in Figure 5d.

Ensemble results are compared with those of quantum mechanics (theoretical values) obtained by using the propagator

$$K^{(BX)}(x, t|x_0) = \sum_{n=1}^{\infty} \exp(-i\pi E_n t)\, \psi^{(n)}(x_0)\psi^{(n)}(x), \tag{119}$$

where $E_n = \frac{n^2}{8a^2}$ and the stationary states are

$$\psi^{(n)}(x) = \sqrt{\frac{1}{a}} \cdot \begin{cases} \cos\dfrac{n\pi x}{2a}, & n = 1,3,5,\ldots \\ \sin\dfrac{n\pi x}{2a}, & n = 2,4,6,\ldots \end{cases} \tag{120}$$

Propagator (119) is equivalent to [35]

$$K^{(BX)}(x, t|x_0) = \lim_{N_s \to \infty} \frac{1}{\sqrt{2\iota t}} \cdot \sum_{l=-N_s}^{N_s} (-1)^l \exp\left(\iota\pi \frac{(x - 2la - (-1)^l x_0)^2}{2t}\right), \tag{121}$$

that is, the propagator of a free particle with $2N_s + 1$ equiprobable sources ("virtual sources"), equally separated by $2a \pm 2x_0$, where $N_s \approx t/(2a)$, and with alternating source phase. Results for single-source and stationary-state preparations are shown.

### 9.4.1 Single source

In this case, $x_0 = \{x_0\}$, $P_0 = \{1\}$. The theoretically expected pdf for the position is

$$\rho(x;t) = \frac{1}{2a} \sum_{m=0}^{a-1} \delta(x - \hat{x}^{(m)}). \tag{122}$$

where $\hat{x}^{(m)} = 2a/\pi \arcsin(\sin \pi \hat{v}^{(m)} t/(2a))$ and $\hat{v}^{(m)} = \pm(2m+1)/(2a)$. The latter expression differs from that of a free particle from multiple sources because of the source phase.

Figure 19 shows the distribution of the quantum momentum after $N_t = 500$ iterations with $a = 10$ and $N_p = 500$. Momentum tends to assume definite values $\pm 1/(2a), \pm 3/(2a), ...$, i.e., values $\hat{v}^{(m)}$, although the last values tend to merge (only $a - 1$ values are visible instead of the expected $a$) due to the limited computational resources.

On the other hand, to see the correct interference pattern building in the spatial domain, one would require large $a$ and $t$, which makes large $N_s$ and consequently too long simulations to be afforded.

### 9.4.2 Stationary state

In this case, the source probability and phase are set to represent QM initial states (120). Since these are stationary states, the theoretically expected pdf is obtained as

$$\rho(x;t) = P_0^{(x)} = \frac{1}{a} \cdot \begin{cases} \cos^2\left(\frac{n\pi x}{2a}\right), & n = 1,3,5,... \\ \sin^2\left(\frac{n\pi x}{2a}\right), & n = 2,4,6,... \end{cases}, \tag{123}$$

that is, a function with peaks at $\hat{x}^{(m)} = \pm a/n(2m + m_0)$, with $m_0 = 0$ for $n$ odd and $m_0 = 1$ for $n$ even. Correspondingly, momentum distribution peaks at $\hat{v} = \pm n/(2a), n > 0$.

Figure 24 and Figure 25 show the distribution of the quantum momentum after $N_t = 500$ iterations and for $N_p = 5000$, with $a = 10$ and $n = 3, 5$, respectively. Momentum clearly tends to the theoretically allowed values $\pm n/(2a)$.

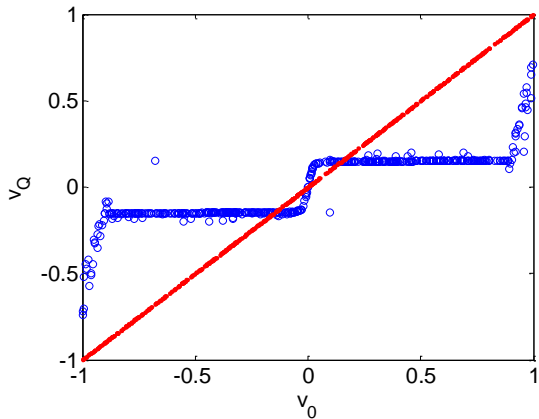

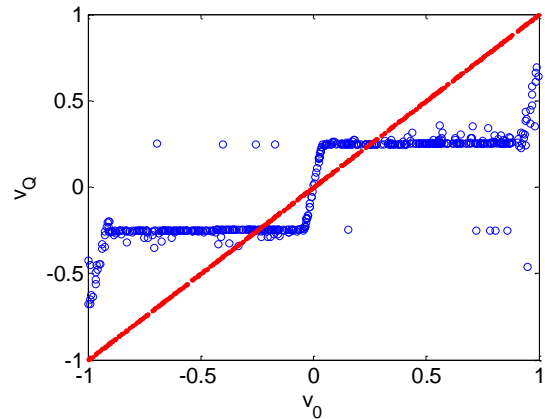

Figure 24: Quantum momentum distribution for $N_p = 500, t = 500$ as a function of source momentum

Figure 25: Quantum momentum distribution for $N_p = 500, t = 500$ as a function of source momentum

(particle in a box, stationary state, $a = 10$, $n = 3$).     (particle in a box, stationary state, $a = 10$, $n = 5$).

## 9.5 Delta potential

This scenario is defined by the amplitude $\lambda$ of the Delta-type function (centred at $x = 0$) that describes the potential. This function is represented in the proposed model by setting $f(x) = \lambda/\ell$ for $x \in (-\ell, 0]$ and $f(x) = -\lambda/\ell$ for $x \in (-2\ell, -\ell]$, with $\ell$ an arbitrary scale (finite rectangular barrier).

Example trajectories of particles emitted from the same source with the same source momentum are plotted in Figure 5e-f. The trajectories clearly differ from the classical ones as the latter either coincide with a free motion $x = x_0 + v_Q t$, if $\lambda < v_0^2/2$, or with a reflected motion $x = \min(x_0 + v_Q t, -x_0 - v_Q t)$, if $\lambda > v_0^2/2$.

Ensemble results are compared with those of quantum mechanics (theoretical values) obtained by using the propagator

$$
\begin{aligned}
K^{(DP)}(x,t|x_0) &= \frac{1}{\sqrt{2\iota t}} \exp\left(\frac{\iota \pi (x-x_0)^2}{2t}\right) \\
&\quad - \frac{\lambda \pi^2}{\sqrt{2\iota t}} \int_0^\infty du \exp\left(-u\pi^2 \lambda - \frac{\pi(|x|+|x_0|+u)^2}{2\iota t}\right) = \\
&= K^{(FP)}(x,t|y) \\
&\quad - \lambda \pi^2 \int_0^\infty du \exp(-u\pi^2\lambda) K^{(FP)}\bigl(x,t|-x/|x|(|x_0|+u)\bigr)
\end{aligned}
\quad (124)
$$

that is clearly equivalent to that of a free particle with an infinity of "virtual sources" placed at positions $-\text{sign}(x) \cdot (|y| + u)$, $u = 0, \ldots$ Considering $y > 0$, for $x > 0$ these virtual sources are at $-y - u$, while for $x < 0$ they are at $y + u$. In both cases, they correspond to delayed bounces of the particle at the potential Delta.

In the proposed model, the virtual sources arise naturally as a consequence of the random motion around $x = 0$. In fact, particles can emerge from the potential Delta with several momentum values, depending on how much time they have spent in the finite rectangular barrier. With respect to the classical case where a particle bounces on the barrier if its energy is lower than the potential and its momentum is reverted ($v_Q \to -v_Q$), here $v_Q \to (v_Q + f\tau^+ - f\tau^-)$, where $\tau^\pm$ is the time spent in the front, resp., rear side of the barrier. Assuming no quantum forces (see later), this momentum is kept in the absence of external forces and if, after $t$ iterations the particle has reached the node $x$, its overall trajectory is equivalent to that of a particle emitted at $x - v_Q t$, that is, to a virtual source with at $u = -x_0 \pm (x + v_Q t)$. As a consequence, the $u$'s are not integers but rational numbers as the $v_Q$'s are.

Virtual sources can be lumped at their average value $-y - 1/\lambda\pi^2$ (with a $\pi$ phase) for $x > 0$ and $y + 1/\lambda\pi^2$ (with zero phase) for $x < 0$, such that an approximation of (124) is

$$
K^{(DP)}(x,t|y) \approx K^{(FP)}(x,t|y) - K^{(FP)}\left(x,t|-\text{sign}(x) \cdot \left(|y| + \frac{1}{\lambda\pi^2}\right)\right). \quad (125)
$$

Results for single-source and Gaussian-wave preparations are shown.

### 9.5.1 Single source

In this case, $x_0 = \{x_0\}$, $P_0 = \{1\}$. By using the approximated propagator (125) the theoretically expected pdf is approximated as

$$\rho(x;t) \approx \begin{cases} \dfrac{1}{2t} \cdot \dfrac{(x-x_0)^2}{\lambda^2\pi^2 t^2 + (x-x_0)^2}, & x < 0 \\ \dfrac{1}{2t} \cdot \dfrac{(x+x_0)^2 + 2\lambda^2\pi^2 t^2}{(x+x_0)^2 + \lambda^2\pi^2 t^2}, & x > 0 \end{cases}. \qquad (126)$$

From (126) the transmission ratio is retrieved (for $x_0 \ll t$) as

$$TRA := \int_{-t}^{0} \rho(x;t) dx \approx \frac{1}{2t}\left[x - \lambda\pi t\,\text{atan}\left(\frac{x}{\lambda\pi t}\right)\right]_{-t}^{0} = \frac{1}{2} - \frac{\lambda\pi}{2}\,\text{atan}\left(\frac{1}{\lambda\pi}\right), \qquad (127)$$

and similarly the "reflection ratio" $REF = 1 - TRA$.

Figure 26 shows the reflection coefficient obtained for various values of $\lambda$ by summing the frequency of arrival at nodes $x < 0$ for $x_0 = 10$, $N_p = 5000$, $t = 200$, and $\ell = 15$. For simplicity the quantum forces have been disabled (the number of virtual sources and thus bosons to take into account would dramatically exceed the capabilities of a standard personal computer) as they do not contribute to the net transmission effect.

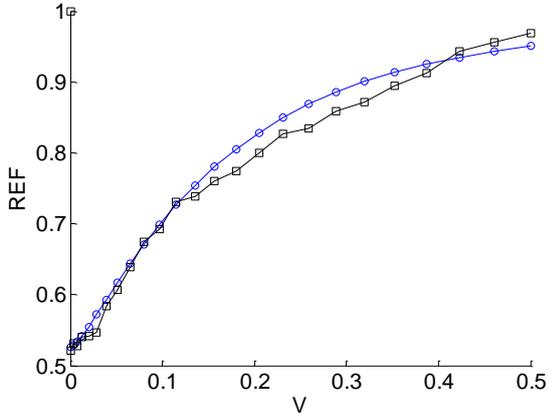
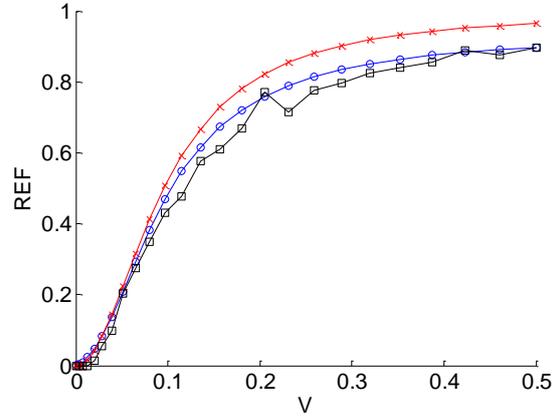

Figure 26: Reflection factor (black) and theoretical value (blue) for $N_p = 5000$, $t = 200$ as a function of potential $\lambda$ (Delta potential, single source, $x_0 = 10$).

Figure 27: Reflection factor (black), theoretical value (blue) and approximation eq. (131) (red) for $N_p = 500$, $t = 3000$ as a function of potential $\lambda$ (Delta potential, Gaussian wave, $a = 10$, $d = 10$, $v_\varphi = -0.3$).

### 9.5.2 Gaussian wave

In this case the sources are prepared as in Sect. 9.1.e. The theoretically expected pdf is found by applying the propagator (125) to the initial state $\psi_0$ shown in that section and is found as $\rho(x;t) = |\psi(x,t)|^2$, with [37]

$$\psi(x,t) = \psi_0(x,t) \cdot \psi_1(x,t), \qquad (128)$$

$$\psi_0(x,t) := \sqrt{\frac{d}{\sqrt{\pi}\mu(t)}} \exp\left[-\frac{(x - a - v_\varphi t)^2}{2\mu(t)} + \iota\pi v_\varphi x - \iota\pi v_\varphi^2 t/2\right], \qquad (129)$$

$$\psi_1(x,t) := \left\{1 - \pi^2\lambda\sqrt{\frac{\pi\mu(t)}{2}}\,\text{erfc}(D(x,t))\exp\left(D(x,t)^2 + \frac{x+|x|}{\mu(t)}(-\iota\pi v_\varphi d^2 - a)\right)\right\}, \qquad (130)$$

where $\mu(t) := d^2 + \iota t/\pi$ and $D(x,t) := [\pi^2\lambda\mu(t) + (a+|x|) + \iota\pi v_\varphi d^2]/\sqrt{2\mu(t)}$.

For large times, an approximation of $\psi(x,t)$ allows to analytically evaluate the transmission coefficient that, if $d^2 v_\varphi^2 \gg 1$ further holds, reads

$$TRA \approx \frac{1}{1 + \left(\frac{\pi\lambda}{v_\varphi}\right)^2}, \tag{131}$$

that is, the plane-wave transmission coefficient.

Figure 27 shows the reflection coefficient obtained for various values of $\lambda$ by summing the frequency of arrival at nodes $x < 0$ for $a = d = 10$, $v_\varphi = -0.3$, $N_p = 500$, $t = 3000$, and $\ell = 2$. For the same reason as indicated above, the quantum forces due to virtual sources have been disabled as they do not contribute significantly to the net transmission effect.

# 10 Two-and three-dimensional scenarios

The scenarios in this section have been simulated using the $M^{**}$ code due to the otherwise prohibitive computational times. The scenarios considered here, namely, (1) free particle, (2) particle on a ring, and (3) particle on a sphere allow observing the emergence of 2-d interference and quantization of angular momentum in the proposed model.

## 10.1 Free 2-d particle

For all the scenarios in this section, $f \equiv 0$. The expected motion is $x_d = x_{0d} + v_{Qd} t$, $d = 1,2$. Ensemble results are compared with those of quantum mechanics (theoretical values) obtained by using the propagator

$$K^{(FP2)}(x_1, x_2, t | x_{01}, x_{02}) = \frac{1}{2\iota t} \exp\left(\frac{\iota \pi (x_1 - x_{01})^2}{2t}\right) \exp\left(\frac{\iota \pi (x_2 - x_{02})^2}{2t}\right) \tag{132}$$

Results for single-source and multiple-source preparations are shown.

### 10.1.1 Single source

In this case, $x_0 = \{(x_{10}, x_{20})\}$, $P_0 = \{1\}$. The theoretically expected pdf is

$$\rho(x;t) = \frac{1}{(2t)^2}. \tag{133}$$

Results are trivially consistent with those of QM and are not presented here.

### 10.1.2 Multiple sources on a line

In this scenario $N_s$ sources are placed in the 2-d lattice at equal distance on a line, $x_0 = \{(ka, kb)\}$, with source probabilities such that $\sum_{k=1}^{N_s} P_0^{(k)} = 1$, and source phase $\epsilon^{(k)} \equiv 0$. In this case $N_s(N_s - 1)$ types of bosons arise, with $\delta^{(ij)} = \{|i-j|a, |i-j|b\}$ and $\epsilon^{(ij)} = 0$. The theoretically expected pdf is thus

$$\rho(x;t) = \frac{1 + \sum_i \sum_{i \neq i} \sqrt{P_0^{(i)} P_0^{(j)}} \cos\left(\pi \delta_1^{(ij)} \frac{x_1 - \frac{a(i+j)}{2}}{t} + \pi \delta_2^{(ij)} \frac{x_2 - \frac{b(i+j)}{2}}{t}\right)}{2t}. \tag{134}$$

Figure 28 shows the frequency of arrivals after $N_t = 100$ iterations for a source scenario with $a = 2$, $b = 1$, $N_s = 3$, and $P_0^{(j)} = 1/3$, $\forall j$. The frequency clearly tends to the theoretical probability density

(107), shown in the top subplot, with peaks along the lines $ax_1 + bx_2 = 2n\pi$. Figure 29 shows the distribution of the quantity $v_{Q\parallel} := av_{Q1} + bv_{Q2}$, a combination of the two quantum momenta that is parallel to the source line (non-normalized). This quantity clearly peaks at the expected values $\hat{v}_{Q\parallel}^{(n)} = 2n$, with $n = 0, \pm 1, \ldots$

This result could have been obtained by rotating the lattice in such a way that the source line coincides with the $x_1$ axis, neglecting the $x_2$ dimension and applying the result of Sect. 9.1 (multiple source scenario). The peak momenta would be $\hat{v}_Q^{(n)} = \frac{2n}{\kappa}$, where $\kappa = \sqrt{a^2 + b^2}$ is the distance between sources along the source line. These peak values equal those of the normalized quantity $\kappa v_{Q\parallel}$.

In other terms, the correct behaviour is obtained in the proposed model even without rotating lattice, which is assumed to be fixed. The model is thus invariant with respect to source rotations.

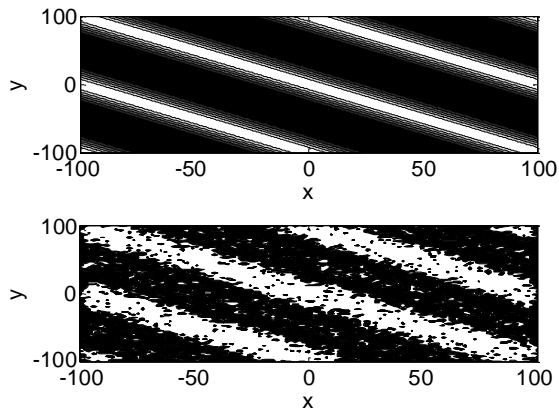
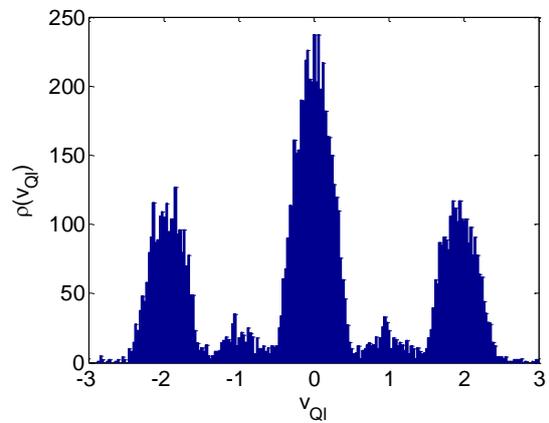

Figure 28: Arrival frequency (down) and theoretical value (top) for $N_p = 10000$, $t = 100$ as a function of position (2D free particle, multiple sources, $a = 2$, $b = 1$, $N_s = 3$).

Figure 29: Distribution of quantum momentum normal to the sources for $N_p = 10000$, $t = 100$ (2D free particle, multiple sources, $a = 2$, $b = 1$, $N_s = 3$).

## 10.2 Particle on a ring (1d periodic motion)

In this scenario the particle is confined on a set of lattice nodes such as $\text{round}[x_1^2 + x_2^2] = r^2$, where $r$ is the ring radius (an integer). With the considerations in Sect. 6.5, the external bosons are such that $a_2 \approx -x_2$ and $a_1 \approx x_1$, while $\kappa = 1/\sqrt{a_1^2 + a_2^2} \approx 1/r$. Consequently, the only relevant combination of momenta is that along the line, $v_{Q\parallel} = \frac{x_2}{r} v_{Q1} - \frac{x_1}{r} v_{Q2}$, which is recognized to be the peripheral momentum $v_{Q\phi}$. Similarly, the only relevant span term is the distance along the ring $\ell_\parallel = \frac{1}{r}\Sigma(-x_2 v_1 + x_1 v_2)$, which is found to equal the circumference arc $\ell_\phi := r(\phi - \phi_0)$, where $\phi$ is the polar angle measured clockwise from the $(0, r)$ point.

This situation is equivalent to a 1d free periodic motion described by the polar coordinate $\phi$. Quantum forces arise even in the presence of a single source because a given node on the ring can be visited by particles having looped through the ring a different number of times, thus with a peripheral span $\ell_\phi$ that may differ for a multiple of the ring circumference. The expected motion is
$$\phi(t) = \text{asin}\left(\sin\frac{v_{Q\phi}}{r}t\right).$$

Ensemble results are compared with those of quantum mechanics (theoretical values) obtained by using the propagator

$$K^{(PR)}(\phi,t|\phi_0) = \sum_{m=-\infty}^{m=\infty} \left(\frac{\pi r^2}{2\pi\iota t}\right)^{\frac{1}{2}} \exp\frac{\iota\pi r^2(\phi-\phi_0-2\pi m)^2}{2t}. \tag{135}$$

This propagator is equivalent to that of a 1d free motion with an infinity of equally-probable virtual sources separated by $2\pi r$. Results for single-source, plane-wave and stationary-state preparations are shown.

### 10.2.1 Single source

In this case, according to section 9.1.d (free particle with multiple sources separated by $2\pi r$), the pdf of the momentum has Dirac peaks at $\hat{v}_\phi^{(m)} = \frac{m}{\pi r}$, with $m = 0, \pm 1, \pm 2 \ldots$ Consequently, the kinetic energy has peak values $\hat{E}^{(m)} = \frac{m^2}{2\pi^2 r^2}$ and the angular momentum has peak values $\hat{J}^{(m)} = \frac{m}{\pi}$, in agreement with QM predictions in lattice units. Results are obvious and not shown.

### 10.2.2 Plane wave

A plane wave state is obtained by placing $[2\pi r] + 1$ equiprobable sources separated by one lattice node ($\phi_0 = \{k\}$, $k \in [-\lfloor\pi r\rfloor, \lceil\pi r\rceil]/r$, $P_0^{(k)} = 1/([2\pi r]+1)$), and providing these sources with a phase $\epsilon^{(k)} = mk$. The theoretically expected pdf is found by applying the propagator (135) to the initial state $\psi_0 = 1/\sqrt{[2\pi r]+1} \cdot e^{\iota m\phi}$ and is

$$\rho(\boldsymbol{\phi};t) \approx \frac{1}{[2\pi r]+1}. \tag{136}$$

Figure 30 shows the frequency of arrival after $N_t = 1000$ iterations with $r = 10$ and $m = 4$. Figure 31 shows the distribution of the quantum momentum. The frequency clearly tends to the theoretical pdf, while the momentum clearly tends to the theoretically allowed value $m/(\pi r)$.

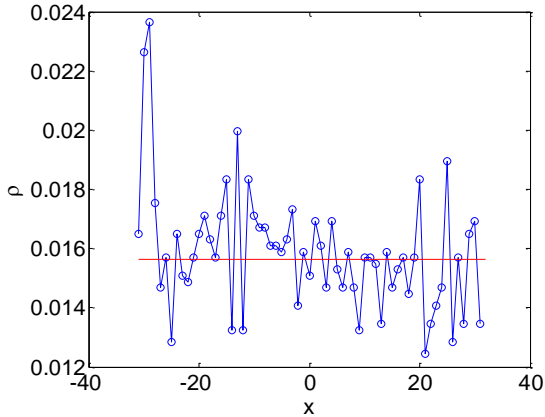
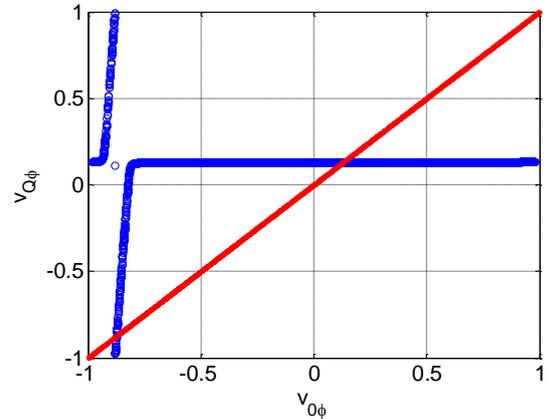

Figure 30: Arrival frequency (blue) and theoretical value (red) for $N_p = 5000$, $t = 1000$ as a function of position (particle on a ring, plane wave, $r = 10$, $m = 4$).

Figure 31: Quantum momentum distribution for $N_p = 5000$, $t = 1000$ as a function of source momentum (particle on a ring, plane wave, $r = 10$, $m = 4$).

### 10.2.3 Stationary state

In this scenario the source probability and phase are prepared so to represent QM initial states $\psi_0^{(n)}(\phi) = \frac{1}{\sqrt{\pi r}}\sin n\phi$ for $n \in \mathbb{Z}$ odd and $\psi_0^{(n)}(\phi) = \frac{1}{\sqrt{\pi r}}\cos n\phi$ for $n \in \mathbb{Z}$ even. Since these are stationary states, the theoretically expected pdf is obtained as $\rho(\boldsymbol{\phi};t) = |\psi_0(\boldsymbol{\phi})|^2$. This function has peaks at $\hat{x}^{(m)} = \pm \frac{1+2m}{2n}$, $m = 0, \ldots, n-1$ for $n$ odd and $\hat{x}^{(m)} = \pm \frac{m}{n}$, $m = 0, \ldots, n$ for $n$ even. The theoretical momentum pdf is a Dirac delta function $\rho_{v_{Q\phi}}(v_{Q\phi}) = \delta\left(v_{Q\phi} \pm \frac{n}{\pi r}\right)$ with two symmetric eigenvalues.

Figure 32 shows the frequency of arrival after $N_t = 1000$ iterations with $r = 10$, $n = 1$. Figure 33 shows the distribution of the peripheral momentum. The frequency clearly tends to the theoretical pdf, while the momentum tends to the theoretically allowed values $\pm \frac{n}{\pi r}$.

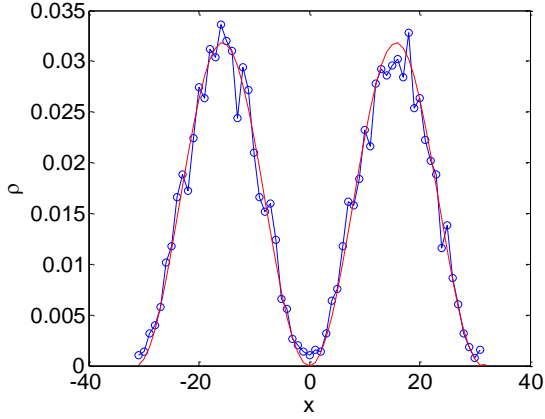
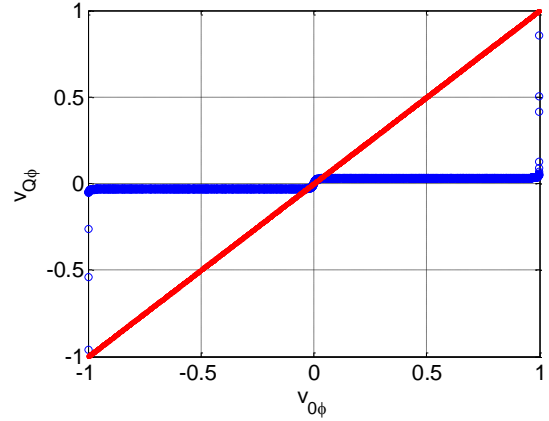

Figure 32: Arrival frequency (blue) and theoretical value (red) for $N_p = 50000$, $t = 1000$ as a function of position (particle on a ring, stationary state, $r = 10$, $n = 1$).

Figure 33: Quantum momentum distribution for $N_p = 50000$, $t = 1000$ as a function of source momentum (particle on a ring, stationary state, $r = 10$, $n = 1$).

## 10.3 Particle on a sphere

In this scenario the particle is confined on a set of lattice nodes such as $\text{round}[x_1^2 + x_2^2 + x_3^2] = r^2$, where $r$ is the sphere radius (an integer). With the considerations in Sect. 6.5, the external bosons are such that $a_1 \approx x_1$, $a_2 \approx x_2$ and $a_3 \approx -x_3$, while $\kappa = 1/\sqrt{a_1^2 + a_2^2 + a_3^2} \approx 1/r$. Consequently, there are two relevant combinations of momenta that are both parallel to the surface and orthogonal to each other. The natural choice is $v_{Q\theta} = -rv_{Q3}/\sqrt{x_1^2 + x_2^2}$, and $v_{Q\phi} = (-x_2 v_{Q1} + x_1 v_{Q2})\sqrt{x_1^2 + x_2^2}/r^2$, which are recognized to be the momentum along meridian arcs and circle of latitude arcs, respectively. Similarly, the only relevant span terms are $\ell_\theta = r \sum v_3/\sqrt{x_1^2 + x_2^2}$ and $\ell_\phi = \frac{1}{r^2} \sum (-x_2 v_1 + x_1 v_2)\sqrt{x_1^2 + x_2^2}$ which are found to equal the arcs $r(\vartheta - \vartheta_0)$ and $r(\varphi - \varphi_0)$, respectively, where $\vartheta$ and $\varphi$ are the usual colatitude and azimuthal angles.

We shall assume, without loss of generality, $\vartheta_0 = \pi/2$ and $\varphi_0 = 0$. The average motion is known from classical mechanics to be

$$\begin{cases} \vartheta(t) = \cos^{-1}(-A \sin(\Omega t)) \\ \varphi(t) = -\frac{\pi}{2} + \tan^{-1}\left(-\frac{\Omega \cot(\Omega t)}{v_{Q\phi}}\right) \end{cases}, \tag{137}$$

which are the equations of a great circle passing through the source, where $\Omega := \text{sign}(v_{Q\theta}) \cdot \frac{1}{r}\sqrt{v_{Q\theta}^2 + v_{Q\phi}^2}$ is the rotational momentum and $A := \sqrt{1 - \frac{v_{Q\phi}^2}{\Omega^2 r^2}}$.

Ensemble results are compared with those of quantum mechanics (theoretical values) obtained by using the propagator

$$K^{(PS)}(\vartheta,\varphi,t|\vartheta_0,\varphi_0) = \sum_{\ell=0}^{\infty} \frac{2\ell+1}{4\pi} e^{-\iota \frac{\ell(\ell+1)}{2\pi r^2} t} P_\ell(\cos(\alpha)), \tag{138}$$

where $\alpha = \text{acos}(\cos\vartheta \cos\vartheta_0 + \sin\vartheta \sin\vartheta_0 \cos(\varphi-\varphi_0))$ is the central angle between the points $(\vartheta,\varphi)$ and $(\vartheta_0,\varphi_0)$. Propagator (138) is approximated [38] as

$$K^{(PS)}(\vartheta,\varphi,t|\vartheta_0,\varphi_0) \approx \mathcal{N}\left(\frac{r^2}{2\iota t}\right)\sqrt{\frac{\alpha + 2\pi n_{loo}}{\sin\alpha}} \exp\left(\frac{\iota \pi r^2 \alpha^2}{2t}\right), \tag{139}$$

where $\mathcal{N}$ is the normalization factor, and can be expressed in terms of $\vartheta$ and $\varphi$ by using the relationship $\Omega t = \alpha + 2n_{loo}\pi := \alpha + 2\pi\left\lfloor \frac{\Omega t}{2\pi} \right\rfloor$, where $n_{loo} = 0, \ldots, \left\lfloor \frac{t}{2\pi} \right\rfloor$ represents the number of times that the particle has looped through the great circle ($\lfloor \cdot \rfloor$ is the "floor" rounding function). By summing over $n_{loo}$ and evaluating the normalization factor such that $\int_0^{2\pi}\int_0^\pi |K^{(PS)}(\vartheta,\varphi)|^2 \sin\vartheta \, d\vartheta \, d\varphi = 1$, the propagator is found to be

$$\begin{aligned}K^{(PS)}(\vartheta,\varphi,t|\vartheta_0,\varphi_0) &\approx \sqrt{\frac{2}{\pi}}\left(\frac{1}{\iota t}\right)\sqrt{\frac{\alpha + t^2/(4\pi)}{\sin\alpha}} \exp\left(\frac{\iota \pi r^2 \alpha^2}{2t}\right) \\ &\approx \frac{1}{\sqrt{2}}\left(\frac{1}{\iota \pi}\right)\sqrt{\frac{1}{\sin\alpha}} \exp\left(\frac{\iota \pi r^2 \alpha^2}{2t}\right), \end{aligned} \tag{140}$$

### 10.3.1 Single source

From (137), a given node on the spherical surface can be visited by particles having the same latitudinal span $\ell_\theta$ but the azimuthal span $\ell_\phi$ may differ by a multiple of $2\pi r$. According to section 9.1 (free particle with multiple sources separated by $2\pi r$), the azimuthal momentum pdf ultimately peaks at $\hat{v}_\phi^{(m)} = \frac{m}{\pi r}$, with $m = 0, \pm 1, \pm 2 \ldots$ However, for large $r$, this pdf is reasonably continuous.

The joint pdf of the two position coordinates can be evaluated analytically from the definition (81) and the average motion (137) as

$$\rho(\vartheta,\varphi;t) \propto \frac{\Omega \sin\vartheta}{t \sin\Omega t} \tag{141}$$

and coincides with the theoretically expected pdf $|K^{(PS)}|^2 \sin\vartheta$ after having summed over $n_{loo}$.

### 10.3.2 Stationary states

In this scenario the source probability and phase are prepared so to represent QM initial states $\psi^{\ell m} := Y_{\ell m}(\vartheta,\varphi)$, where $Y_{\ell m}$ are spherical harmonic functions. Since these are stationary states, the theoretically expected pdf is obtained as

$$\rho(\boldsymbol{x};t) \approx P_0^{(x)} \propto \frac{2\ell+1}{4\pi} \frac{(\ell-m)!}{(\ell+m)!}(P_\ell^m(\cos\vartheta))^2. \tag{142}$$

Each state implies a source distribution with $\ell + 1 - m$ peaks separated by $\Delta\hat{\theta}^{(\ell m)} \approx \frac{\pi}{\sqrt{\ell(\ell+1)-m^2/\sin^2\hat{\theta}}}$, while each $\phi$ node is equally probable. This leads to a latitudinal momentum distribution that, in accordance to the scenario A.1.c (free 1d particle, multiple sources), has peaks for $\hat{v}_{Q\theta}^{(\ell m)} = \pm \frac{\sqrt{\ell(\ell+1) - m^2/\sin^2\hat{\theta}}}{\pi r}$. On the other hand, as scenario B.1.b (particle on a ring of radius $r_r = r\sin\vartheta$) has shown, the azimuthal momentum distribution peaks at $\hat{v}_{Q\phi}^{(\ell m)} = r\hat{\dot{\phi}}^{(ring)} = \frac{r\dot{v}_{\phi}^{(ring)}}{r_r} = \frac{m}{\pi r \sin^2\hat{\theta}}$. Consequently, the energy peaks at $\hat{E} = \frac{1}{2}\left(\left(\hat{v}_Q^{(\ell m)}\right)^2 + \left(\hat{v}_\phi^{(\ell m)}\right)^2 \sin^2\hat{\theta}\right) = \frac{\ell(\ell+1)}{2\pi^2 r^2}$, while the z-axis component of the angular momentum ($J_z = rv_{Q\phi}\sin^2\vartheta$) peaks at $\hat{J}_z^{(\ell m)} = \frac{m}{\pi}$, as predicted by QM.

These results have been used to accelerate the simulations that, for $[\pi r] \times [2\pi r]$ sources would otherwise imply a dramatically high number of bosons. In fact, the sources have been placed on a single great arc and the particle evolution has been computed only for that arc. Figure 34 shows the frequency of arrival after $N_t = 100$ iterations with $r = 100$, $\ell = 4$, $m = 0$. Figure 35 shows the distribution of the latitudinal momentum. The frequency clearly tends to the theoretical pdf, while the momentum tends to the theoretically allowed values $\pm \frac{\sqrt{\ell(\ell+1)}}{\pi r}$ ($\pm 0.0142$ in this case).

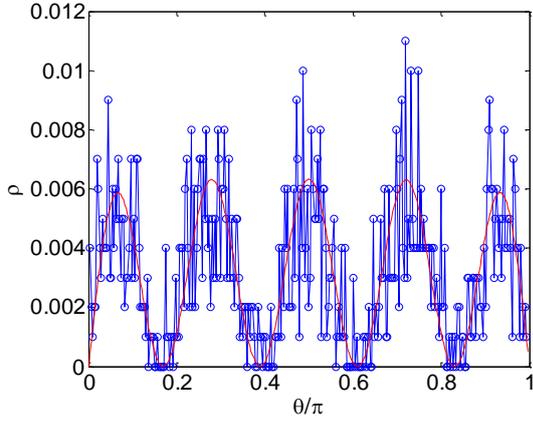
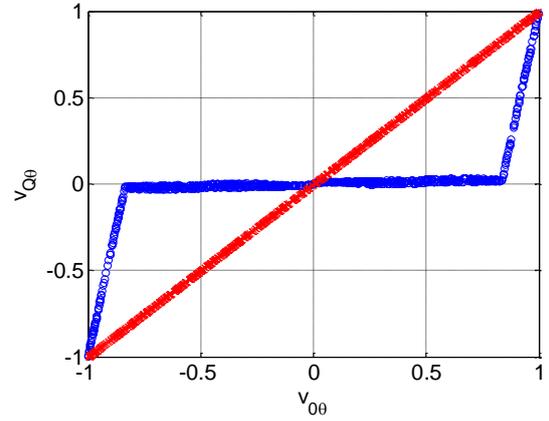

Figure 34: Arrival frequency (blue) and theoretical value (red) for $N_p = 1000$, $t = 100$ as a function of position (particle on a sphere, stationary state, $\ell = 4$, $m = 0$).

Figure 35: Latitudinal quantum momentum distribution for $N_p = 1000$, $t = 100$ as a function of source momentum (particle on a sphere, stationary state, $\ell = 4$, $m = 0$).

## 11 Two-particles systems

### 11.1 Entangled particles

Instead of reproducing a realistic two-particle interferometer containing at least a pair of beam splitters [40], which are systems with internal characteristics that are complex to model [41], we choose a simpler system yet able to test Bell's inequalities. The interferometer used here is a two-slit interferometer.

#### 11.1.1 Double-slit interferometer

Consider the non-entangled case first, where single particles are emitted. The considered interferometer is equivalent to the double-source preparation of Sect. 9.1.2 with $P_0^{(\pm D)} = 1/2$ and the additional requirements that $\epsilon^{(D)} = \alpha$ while $\epsilon^{(-D)} = 0$, so that particles experience a phase difference $\alpha$. Consequently, the theoretically expected pdf is

$$\rho(x;t) = \frac{1 + \cos\left(2\pi D \frac{x}{t} - \pi\alpha\right)}{2t}. \tag{143}$$

If two "detectors" are placed at a time $t$ at positions $x_\pm = \pm t/4D$, the pdf of hitting such detectors is

$$\rho(x_\pm;t) = \frac{1 \pm \sin(\pi\alpha)}{2t} \tag{144}$$

that is, in formal agreement with Malus' law and QM.

The proposed model naturally reproduces the theoretical prediction as shown in Sect. 9.1.2.

In the entangled preparation case, the emission rules of Sect. 7.1 are applied, with sources $\{a,b\} = \{D,-D\}$. Each particle of an entangled pair has a motion that is fully independent from the other, once the source quantities have been imprinted. To represent an interferometer where phases of different paths can be arbitrarily adjusted, particles I are further prepared according to Sect. 7.1 to experience a phase difference $\epsilon^{(I)} = \alpha/\pi$, while the phase difference for particles II is $\epsilon^{(II)} = \beta/\pi$. Once emitted, particles I and II evolve independently as in Sect. 7.2. Alternatively, this scenario may be thought as if particles I are emitted in the, say, negative $x_2$ direction, and particles II in the positive direction, with unitary momentum along those directions. Detectors are placed at a time $t$ at positions $x_\pm = \pm t/4D$ as in the non-entangled case.

The theoretically expected joint pdf's of arrivals are proportional to the quantities $1 \pm \cos(\pi(\beta - \alpha))$, where the plus sign is for detectors placed at the same positions, and the minus sign for detectors placed at opposite positions. Consequently, the correlation factor is

$$C = \cos(\pi(\beta - \alpha)) \tag{145}$$

and the Bell-CHSH quantity [42, 43] is evaluated as

$$S_{max}(\theta) = 3\cos\theta - \cos 3\theta, \tag{146}$$

a function that violates the Bell-CHSH inequalities for some values of $\theta/\pi = \beta - \alpha = \alpha' - \beta = \beta' - \alpha$.

In the proposed model, the scenario falls under the assumptions for which (90)-(98) apply. In particular, (98) yields the expected pdf's

$$\rho(x_+, x_+; t) = \rho(x_-, x_-; t) = \frac{1}{4t^2}(1 + \cos(\pi(\beta - \alpha))) \tag{147}$$

and

$$\rho(x_+, x_-; t) = \rho(x_-, x_+; t) = \frac{1}{4t^2}(1 - \cos(\pi(\beta - \alpha))). \tag{148}$$

For the correlation coefficient we therefore find the form (145), in perfect agreement with the QM prediction.

Figure 36 shows the four frequencies of arrival after $N_t = 100$ iterations, with emissions at $\pm D = \pm 1$ and, consequently, detectors at $x_\pm = \pm 25$, for 21 different values of $\alpha$, and $\beta = 0$. Figure 37 shows the frequencies of arrivals for $\beta = 0.3$ and varying $\alpha$. Clearly, the patterns tend to (147)-(148) and they would approach the theoretical values even more for larger values of $t$. However, larger times would also require larger numbers of emissions to record a significant number of detection coincidences, as this number scales with the factor $N_p/t^2$, making the simulation time unpractically long.

Figure 38 shows the correlation factor obtained in both settings ($\beta = 0$ and $\beta = 0.3$, varying $\alpha$), which also tends to the theoretical value (145).

The Bell-CHSH quantity is shown in Figure 39. This quantity has been obtained from the definition $S_{max} = 3C(\theta) - C(3\theta)$, by setting $\beta = 0$ and varying $\theta$, performing the simulation first for $\alpha = \theta$ and then for $\alpha = 3\theta$, then computing the correlation factors. The figure clearly shows that, in agreement with the theoretical values, the computed $S_{max}$ exceeds the Bell limit value of 2 for some values of $\theta$. In other terms, despite being a local and realistic model, the proposed model is able to violate the Bell-CHSH inequalities.

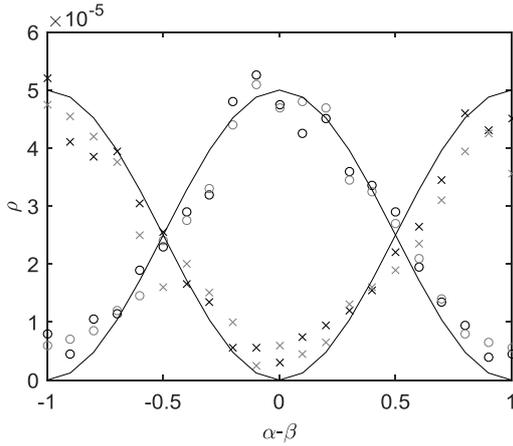

Figure 36: Joint arrival frequencies at pairs of detectors (circles: same position, crosses: opposite positions) and theoretical value (solid) for $N_p = 2 \cdot 10^6$, $t = 100$ as a function of the phase difference with $\beta = 0$ (two-slit interferometer, $D = 1$, $P_0^{(D)} = 0.5$).

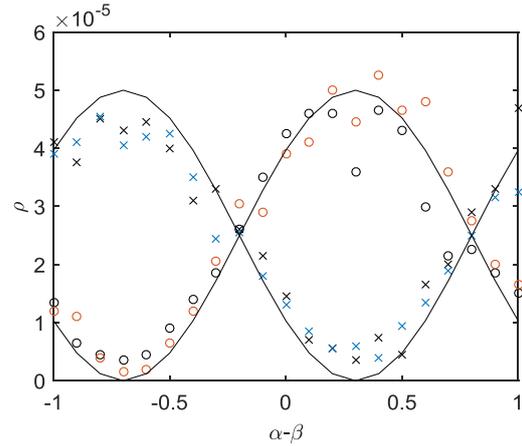

Figure 37: Joint arrival frequencies at pairs of detectors (circles: same position, crosses: opposite positions) and theoretical value (solid) for $N_p = 2 \cdot 10^6$, $t = 100$ as a function of the phase difference with $\beta = 0.3$ (two-slit interferometer, $D = 1$, $P_0^{(D)} = 0.5$).

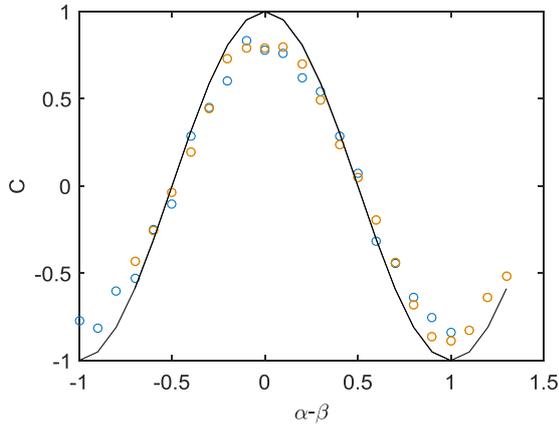

Figure 38: Correlation factor (blue circles: $\beta = 0$, orange circles: $\beta = 0.3$) and theoretical value (black) for $N_p = 2 \cdot 10^6$, $t = 100$ as a function of the phase difference (two-slit interferometer, $D = 1$, $P_0^{(D)} = 0.5$).

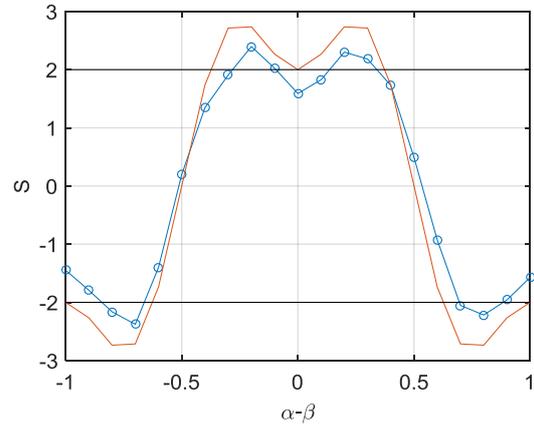

Figure 39: CHSC quantity (circles) and theoretical value (red) for $N_p = 2 \cdot 10^6$, $t = 100$ as a function of the phase difference (two-slit interferometer, $D = 1$, $P_0^{(D)} = 0.5$).

# 12 Discussion

The paper has presented a model aimed at reproducing QM results while assuming realistic, stochastic, and local motion rules of individual particles. QM predictions are indeed retrieved as probability distributions of position, momentum, and related quantities without appealing to the QM mathematical machinery itself.

The model uses a single and relatively simple set of motion rules, assuming that (1) particles evolve in a discrete spacetime lattice; (2) the evolution on the lattice is a random walk; (3) the transition probabilities of the random walk are simple functions of a momentum propensity that results from three mechanisms, namely, (3a) random particle preparation at the sources, (3b) interaction with external "bosons" representing external forces, and (3c) exchange with quantum "bosons" left at the lattice nodes by previously emitted particles; (4) the latter information is progressively built as the lattice nodes are visited by successive emissions, each leaving a kind of footprint.

Despite its compactness, the model has proven sufficient to cover various QM scenarios, from free particle preparations to various external forces (free faller, harmonic oscillator), infinite potential walls (particle in a box), finite potential walls (Delta potential) with quantum tunnelling, constrained motion (free particle on a ring, on a sphere) leading to quantization of energy levels and angular momentum, as well as momentum entanglement.

As the simulations presented have shown, *observables* are described by variables (mostly integer- or rational- valued) that are attributed to the particles or to the lattice. These variables are subject to stochastic preparation at sources and time evolution, leading to their description in probabilistic terms (pdf or pmf). QM *states* are represented by the initial conditions as determined at the sources for the ensemble of particles, i.e., the functions $P_0^{(k)}$ and possibly $\epsilon^{(k)}$. To describe state's dynamics, *Schrödinger equation* is replaced by the microscopic rules of motion, including the action of both external and quantum forces. The distribution of external "bosons" $f(x)$, itself a non-primary, derived quantity, embodies the information that in QM is carried by the *Hamiltonian*. *Born rule* to find the pdf or the pmf of a certain observable is replaced by integral formulas such as (82). Stationary states are retrieved as those states whose pdf does not vary with time, e.g., from the implicit equation (64). Depending on the scenario, quantization of momentum, energy, and angular momentum may arise when their respective pdf's tend to a staircase function with peak values. Quantum tunnelling is also naturally described as the stochastic motion rules allow particles to locally violate classical energy barriers. With coupled source momentum and phase, entangled particles are incorporated in the model that, despite being local and realistic, is thus able to violate the Bell-CHSH inequalities.

Several refinements of the model are still possible. For example, relativistic Newton's second law shall inspire a mechanism to prevent that the momentum propensity becomes larger than unity under the action of persistent forces. Multi-state and many-particle systems are yet to be fully studied, too.

# References

1. Feynman RP, Leighton RB, and Sands M, The Feynman Lectures on Physics, Vol. 3, Addison-Wesley, Reading MA (1965).

2. Santos E, Towards a realistic interpretation of quantum mechanics providing a model of the physical world, Foundations of Science 20(4), 357-386 (2015).

3. Harrigan N and Spekkens RW, Einstein, incompleteness, and the epistemic view of quantum states. Foundations of Physics, 40, 125–157 (2010).

4. Einstein A, Podolsky B, and Rosen N, Can quantum-mechanical description of physical reality be considered complete? Physical Review, 47, 777–780 (1935).

5. D'Espagnat B, Quantum physics and reality. Foundations of Physics, 41, 1703–1716 (2011).

6. Bohm D and Hiley BJ, The Undivided Universe: an ontological interpretation of quantum theory, Routledge, London (1993).

7. Floyd E, Welcher weg? A trajectory representation of a quantum diffraction experiment, Found. Phys., 37(9), 1403–1420 (2007).

8. De la Peña L, Cetto AM, and Brody TA, Hidden variable theories and Bell's inequality, Lett. Nuovo Cimento 5, 177 (1972).

9. Aspect A, Bell's theorem: the naive view of an experimentalist, in : Bertlmann R. and Zeilinger A., Quantum [Un]speakables, Springer, Berlin Heidelberg, p. 119-153 (2002).

10. De Raedt K, de Raedt H, and Michielsen K, A computer program to simulate Einstein–Podolsky–Rosen–Bohm experiments with photons, Computer Physics Communications, 176(11-12), 642-651 (2007).

11. Di Lorenzo A, Beyond Bell's theorem: Admissible hidden-variable models for the spin-singlet, Journal of Physics : Conference Series 442 012046 (2013).

12. Feynman RP and Hibbs AR, Quantum Mechanics and Path Integrals, McGraw-Hill (1965).

13. Ord GN, The Schrödinger and diffusion propagators coexisting on a lattice. J. Phys. A, 29, L123–L128 (1996).

14. Ord GN and Deakin AS, Random walks, continuum limits and Schrödinger's equation. Phys. Rev. A, 54, 3772–3778 (1996).

15. Ord GN, Schrödinger's Equation and Classical Brownian Motion, Fortschr. Phys. 46, 6–8, 889–896 (1998).

16. Janaswamy R, Transitional probabilities for the 4-state random walk on a lattice, J. Phys. A: Math. Theor. 41, 1–11 (2008).

17. Badiali JP, Entropy, time-irreversibility and the Schrödinger equation in a primarily discrete spacetime, J. Phys. A: Math. Gen. 38(13), 2835–2848 (2005).

18. Nelson E, Quantum Fluctuations. Princeton University Press (1985).

19. Nottale L, Fractal Space-Time and Microphysics. World Scientific (1996).

20. Nagasawa M, Schrödinger equations and diffusion theory. Birkhauser Verlag (1993).

21. El Naschie MS, A note on quantum mechanics, diffusional interference, and informions. Chaos, Solitons & Fractals, 5(5), 881–884 (1995).

22. Fritsche L and Haugk M, A new look at the derivation of the Schrödinger equation from Newtonian mechanics, Ann. Phys. (Leipzig) 12(6), 371–403 (2003).

23. Carroll R, Remarks on the Schrödinger equation, Inter. Jour. Evolution Equations 1, 23–56 (2005).

24. Grössing G., Sub-quantum thermodynamics as a basis of emergent quantum mechanics, Entropy 12, 1975–2044 (2010).

25. Jin F, Yuan S, De Raedt H, Michielsen C, and Miyashita S, Corpuscular model of two-beam interference and double-slit experiments with single photons, J. Phys. Soc. Jpn. 79(7), 074401-1–14 (2010).

26. Michielsen K and De Raedt H, Event-based simulation of quantum physics experiments, Int. J. Modern Physics C 25(8), 1430003 (2014).

27. Khrennikov A and Volovich YI, Discrete time dynamical models and their quantum-like context-dependent properties. J. Modern Optics 51(6/7), 113-114 (2004).

28. Dowker F and Henson J, Spontaneous Collapse Models on a Lattice, J. Stat. Phys. 115(516):1327–1339 (2004).

29. Bialynicki-Birula I, Weyl, Dirac, and Maxwell equations on a lattice as unitary cellular automata, Phys. Rev. D 49(12), 6920–6927 (1994).

30. Couder Y, Bouadoud A, Protière S, and Fort E, Walking droplets: a form of wave–particle duality at macroscopic level?, Europhysics News, 41(1), 14–18 (2010).

31. Sciarretta A., A local-realistic model of quantum mechanics based on a discrete spacetime (extended version), www.researchgate

32. Snyder HS, Quantized space-time, Phys. Rev. 71(1), 38-41 (1947).

33. Finkelstein D, Saller H, and Tang Z, Quantum spacetime, in: P. Pronin, et al. (Eds.), Gravity, Particles and Space Time, World Scientific, Singapore (1996).

34. Sidharth BG, The Thermodynamic Universe: Exploring the Limits of Physics, World Scientific, Singapore (2008).

35. Fulling SA and Güntürk KS, Exploring the propagator of a particle in a box, Am. J. Phys. 71(1) (2003).

36. Ingold GL, Path integrals and their application to dissipative quantum systems, in: Buchleitner A and Hornberger K (eds.), Coherent Evolution in Noisy Environments, vol. 611 of Lecture Notes in Physics, Springer, Berlin Heidelberg (2002).

37. Dodonov VV and Dodonov AV, Transmission of correlated Gaussian packets through a Delta-potential, J. Russ. Laser Res., 35(1), 39-45 (2014).

38. Ghosh A, Samuel J, and Sinha S, A "Gaussian" for diffusion on the sphere, EPL 98, 30003 (2012).

39. Di Lorenzo A, Beyond Bell's theorem: Admissible hidden-variable models for the spin-singlet, J. of Physics: Conference Series 442 (2013) 012046.

40. Bernstein HJ, Greenberger DM, Horne MA, Zeilinger A, Bell theorem without inequalities for two spinless particles, Phys. Rev. A, 47(1) (1993).

41. Hénault F, Quantum physics and the beam splitter mystery, Proceedings of the SPIE vol. 9570, no. 95700Q (2015).

42. Aspect A, Bell's theorem: The naive view of an experimentalist, in: Quantum [Un]speakables, Springer, Berlin-Heidelberg (2002).

43. Clauser JF, Horne MA, Shimony A, Holt RA, Proposed experiment to test local hidden-variable theories, Phys. Rev. Lett. 23, 880 (1969).

## Acknowledgments

I would like to thank H. de Raedt for very useful discussions.